\documentclass{article}

    \PassOptionsToPackage{numbers, compress}{natbib}
\usepackage{xspace}

\newcommand{\tperd}{\textsf{T/\$}\xspace}

\usepackage[preprint]{neurips_2024}

\usepackage[utf8]{inputenc} % allow utf-8 input
\usepackage[T1]{fontenc}    % use 8-bit T1 fonts
\usepackage{hyperref}       % hyperlinks
\usepackage{url}            % simple URL typesetting
\usepackage{booktabs}       % professional-quality tables
\usepackage{amsfonts}       % blackboard math symbols
\usepackage{nicefrac}       % compact symbols for 1/2, etc.
\usepackage{microtype}      % microtypography
\usepackage{xcolor}         % colors
\usepackage{graphicx}
\usepackage{subcaption}
\usepackage{makecell}
\usepackage{enumitem}
\usepackage{wrapfig}

\usepackage{tikz}
\newcommand{\greycircle}[1]{%
  \tikz[baseline=(char.base)]{
    \node[shape=circle, fill=gray!50, inner sep=0.25pt] (char) {\textbf{\texttt{#1}}};}%
}

\title{Mélange: Cost Efficient Large Language Model Serving by Exploiting GPU Heterogeneity}

\author{
  Tyler Griggs\thanks{Equal contribution} \\
  UC Berkeley \\
  \And
  Xiaoxuan Liu\footnotemark[1] \\
  UC Berkeley \\
  \And
  Jiaxiang Yu \\
  National University of Singapore \\
  \And
  Doyoung Kim \\
  UC Berkeley \\
  \And
  Wei-Lin Chiang \\
  UC Berkeley \\
  \And
  Alvin Cheung \\
  UC Berkeley \\
  \And
  Ion Stoica \\
  UC Berkeley \\
}

\begin{document}

\maketitle

\begin{abstract}
Large language models (LLMs) are increasingly integrated into many online services, yet they remain cost-prohibitive to deploy due to the requirement of expensive GPU instances. Prior work has addressed the high cost of LLM serving by improving the inference engine, but less attention has been given to selecting the most cost-efficient GPU type(s) for a specific LLM service. There is a large and growing landscape of GPU types and, within these options, higher cost does not always lead to increased performance. 
Instead, through a comprehensive investigation, we find that three key LLM service characteristics (request size, request rate, SLO) strongly influence GPU cost efficiency, and differing GPU types are most cost efficient for differing LLM service settings. As a result, the most cost-efficient allocation for a given service is typically a \textit{mix} of heterogeneous GPU types.
Based on this analysis, we introduce Mélange, a GPU allocation framework that navigates these diverse LLM service characteristics and heterogeneous GPU option space to automatically and efficiently derive the minimal-cost GPU allocation for a given LLM service. We formulate the GPU allocation task as a cost-aware bin packing problem where GPUs are bins and items are slices of the service workload. Our formulation's constraints account for a service's unique characteristics, allowing Mélange to be \textit{flexible} to support diverse service settings and \textit{heterogeneity-aware} to adapt the GPU allocation to a specific service. 
Compared to using only a single GPU type,
Mélange reduces deployment costs by up to 77\% in conversational settings, 33\% in document-based settings, and 51\% in a mixed setting.
\end{abstract}

\section{Introduction}
\label{sec:introduction}
Large language models (LLMs)~\cite{openai2023gpt,touvron2023llama,touvron2023llama2} are increasingly integrated into many online services, including search engines~\citep{google-ai-search, bing-ai-search}, chatbots~\citep{chatgpt}, and virtual assistants~\citep{copilot, wu2023autogen, wu2023empirical}. 
These services are often hosted by deploying models on cloud resources.
However, deploying LLMs is expensive. The substantial size and computational demands of LLMs require the use of costly hardware accelerators, typically GPUs\footnote{For brevity, we use ``accelerator'' and ``GPU'' interchangeably in this work.}
For example, serving Llama2-70b at BF16 precision requires 2 NVIDIA A100-80GB GPUs, which costs over $\$5,200$ per month in on-demand rental costs on major cloud platforms.

Prior work~\cite{dao2022flashattention, jin2024s, zhang2024h2o, zheng2024response, zhu2024towards} addresses the high cost of LLM serving by focusing on inference throughput, but less attention has been given to selecting the most cost-efficient GPU type(s) for a specific LLM service. The large and growing landscape of hardware accelerators — ranging from NVIDIA GPUs~\citep{nvidia-site} and AMD GPUs~\citep{databricksTrainingLLMs} to Google TPUs~\citep{jouppi2017datacenter}, CPUs~\citep{luo2022srifty}, and others~\citep{amazonAcceleratorTrainium} — offers a wide array of choices with varying performance specifications and on-demand cloud costs.
Within these hardware options, higher cost does not always lead to increased performance. To investigate this phenomenon further, we examine GPU \textit{cost efficiency}, defined based on common pricing models~\citep{chatgpt} as the number of input and output tokens processed per dollar cost (\textbf{\tperd}) of on-demand cloud GPUs. We find that GPU cost efficiency is determined by three key LLM service characteristics:

\begin{enumerate}[nosep,leftmargin=1.5em,labelwidth=*,align=left]
\item \textbf{Request Size:} 
An LLM request's size is made up of its input and output token lengths.
For small request sizes, lower-end GPUs generally produce greater \tperd than high-end GPUs. 
\item \textbf{Request Rate:} To maximize utilization, provisioned GPU capacity should align with request volume. 
At low request rates, services can reduce costs by right-sizing from expensive high-end GPUs to cheap low-end GPUs. Further, leveraging a \textit{mix} of GPU types facilitates finer-grained resource scaling to better match request volume.
\item \textbf{Service-level Objective:} 
Services typically establish latency SLOs to ensure service quality.
Because low-end GPUs generally incur higher latency than high-end GPUs,
high-end GPUs are required for stringent SLOs while low-end GPUs can reduce costs in loose-SLO settings.  
\end{enumerate}

Consider a GPU allocation strategy that integrates each of the three observations above: high-cost A100 GPUs handle large requests and meet stringent SLOs, but lower-cost A10G GPUs serve smaller requests (\textbf{1}) and looser SLOs (\textbf{3}) at higher \tperd. Then, during periods of low service activity, the service right-sizes to the even-cheaper L4 GPU to maintain service availability at lowest cost (\textbf{2}).
Consequently, we find that GPU \textit{heterogeneity} presents opportunities for increasing GPU cost efficiency, but such opportunities are highly dependent on LLM service characteristics.
The key challenge, then, is creating a GPU allocation framework that can navigate the diversity of LLM services (request sizes, request rates, latency SLOs) and GPU types to find the optimal GPU allocation.

\begin{figure}[h]
    \centering
    \includegraphics[width=0.7\linewidth]{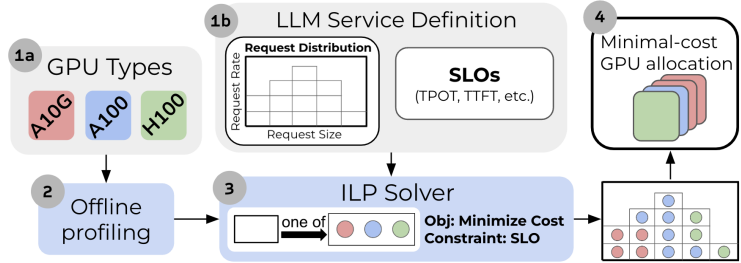}
    \caption{\textbf{Mélange framework.}}
    \label{fig:melange-diagram}
    % \vspace{-1.1em}
\end{figure}

We present \textit{Mélange}\footnote{Mélange is French for ``mixture''} (\autoref{fig:melange-diagram}), a GPU allocation framework that derives the minimal-cost GPU allocation for a given LLM service. In Mélange, each GPU type (\greycircle{1a}) passes through a one-time offline profiling step (\greycircle{2}) to measure GPU performance across request sizes and rates. Then, given the profiling results and an LLM service definition (\greycircle{1b}), Mélange's objective is to choose a GPU allocation for the service workload that minimizes cost.
This task is a natural application of the cost-aware bin packing problem, where bins are GPUs and items are slices of the workload. We formulate the problem as an integer linear program (ILP) and efficiently solve with an off-the-shelf solver (\greycircle{3}). Upon solution, Mélange produces the GPU allocation that can serve the LLM service at minimal cost while adhering to the service SLO (\greycircle{4}). 

Mélange's strength stems from two key properties. First, it is \textit{heterogeneity-aware}. Our analysis shows that request size, request rate, and SLOs jointly impact cost efficiency, but their impacts differ for each GPU type. Mélange's profiling and ILP formulation account for each of these dimensions, enabling efficient navigation of heterogeneous GPU types given a service specification.
Second, Mélange is \textit{flexible}. The inputs (\greycircle{1a}, \greycircle{1b}) can be flexibly modified to include new generations of GPUs or alternative definitions of SLO, ensuring Mélange is effective for diverse services.
Further, to the best of our knowledge, Mélange is the first GPU allocation framework that utilizes multiple GPU types for LLM serving.
In summary, this paper makes the following contributions:
\begin{itemize}[topsep=0em, itemsep=0em, leftmargin=1em,rightmargin=0em]
    \item We analyze three key LLM service characteristics and their influence on GPU cost efficiency: request size, request rate, and latency SLO (\S~\ref{sec:observation}). 
    \item We introduce Mélange, an allocation framework that automatically derives the minimal-cost GPU allocation for a given LLM service while satisfying an SLO requirement (\S~\ref{sec:algorithm}).
    \item We evaluate Mélange across four GPU types—NVIDIA L4, A10G, A100, and H100.
    Mélange reduces costs by 9-77\% for short-context tasks (interactive chats), 2-33\% for long-context tasks (document-based), and 4-51\% in mixed-context workloads (\S~\ref{sec:experiment}).
\end{itemize}
\section{Related Work}
\subsection{LLM Inference Optimization}
A significant body of research has focused on optimizing LLM inference efficiency. One stream concentrates on memory optimization, particularly through improved key-value cache reuse~\citep{zheng2023efficiently} and management strategies~\citep{kwon2023efficient}. Another avenue seeks to minimize latency, such as scheduling optimization~\citep{yu2022orca, agrawal2023sarathi, wu2023fast}, speculative decoding~\citep{leviathan2023fast, kim2024speculative}, kernel optimization~\citep{dao2022flashattention, flashinfer} and early exiting~\citep{teerapittayanon2016branchynet, zhou2020bert}. Additional optimizations include quantization~\citep{frantar2022gptq, lin2023awq, xiao2023smoothquant, yao2022zeroquant} and sparsification~\citep{frantar2023sparsegpt, zaheer2020big}. Instead of altering inference logic, our work assumes a fixed inference engine configuration and concentrates on reducing LLM deployment costs by choosing cost-effective GPU instance types.

\subsection{Machine Learning with Cloud Resources}
Recent studies have explored various strategies for reducing the cost of machine learning (ML) inference or training. Several focus on utilizing spot instances \citep{thorpe2023bamboo,harlap2018tributary,zhang2019mark,gunasekaran2022cocktail}, which is complementary to our work.
Other work targets deployment on heterogeneous resources \citep{borzunov2022petals, chaudhary2020balancing, narayanan2020heterogeneity, miao2023sdpipe, miao2022galvatron}, but focuses primarily on model training rather than serving. Also, lveraging serverless instances for inference cost reduction has been examined in~\citep{ali2022optimizing}. 
Nonetheless, these prior work predominantly concentrate on machine learning prior to the advent of LLMs, which we show to have unique characteristics that significantly impact cost efficiency. More recent studies, such as ~\citep{miao2023spotserve, jiang2023hexgen}, focus on LLMs, but they propose strategies for reducing costs via optimal migration plans and parallelism with heterogeneous resources. They do not identify key LLM service characteristics that impact cost efficiency and consider them in GPU deployment, which our work highlights. Another line of work ~\citep{zhong2024distserve, patel2023splitwise} explores splitting LLM inference into its two phases (prefill and decode) and performing the two phases on separate nodes, perhaps with different GPU types. Our work shows that, even within a phase, the best GPU type can change based on LLM service specifications.  
\section{Background}
\label{sec:background}

\subsection{LLM Request Size Variance}
\label{sec:ob-size-variance}
\begin{figure}[!htbp]
    \centering
    \begin{subfigure}[b]{0.25\linewidth}
    \includegraphics[width=1\linewidth]{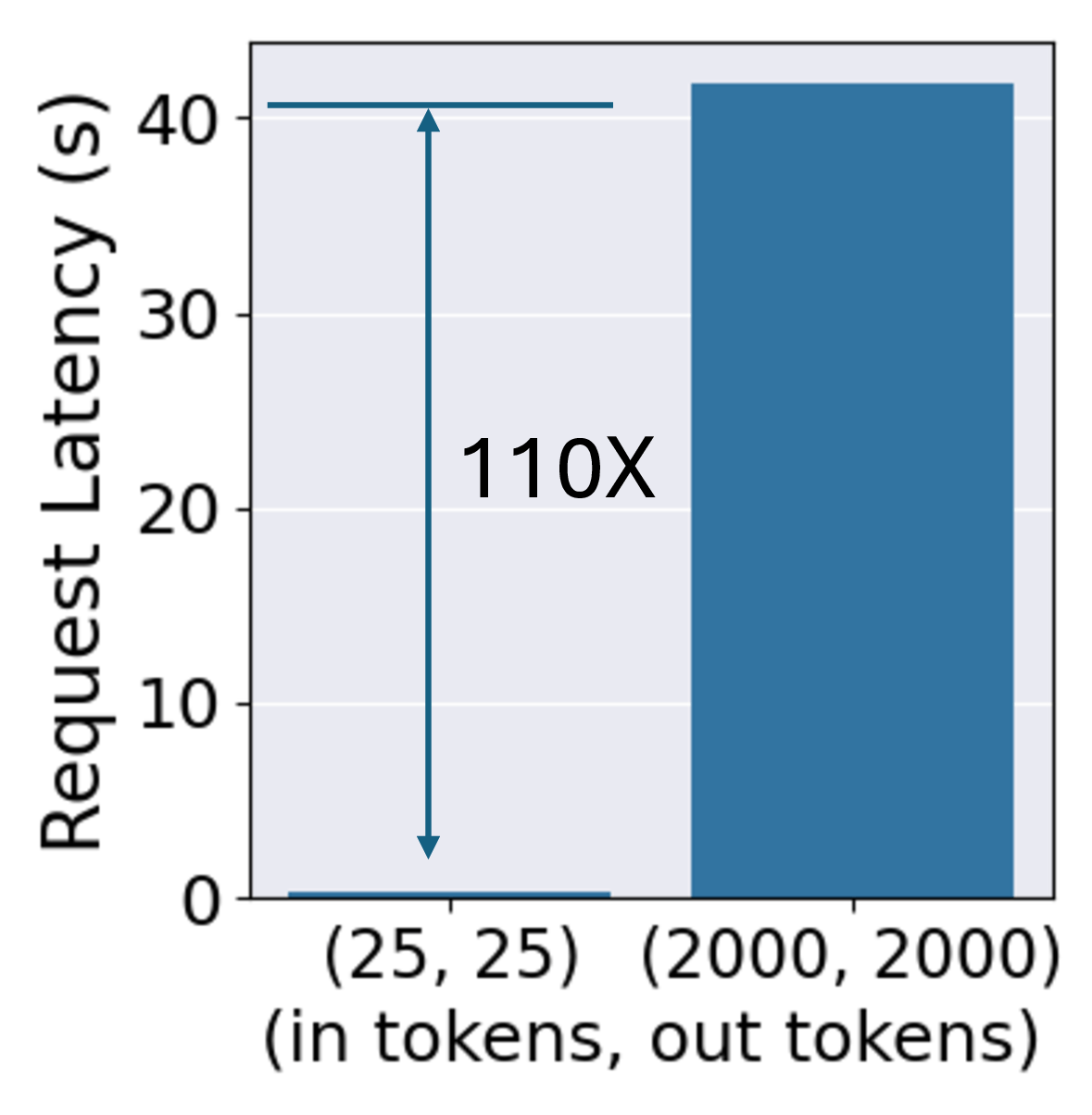}
    \caption{LLaMA-7B}
    \end{subfigure}
    \begin{subfigure}[b]{0.25\linewidth}
    \includegraphics[width=1\linewidth]{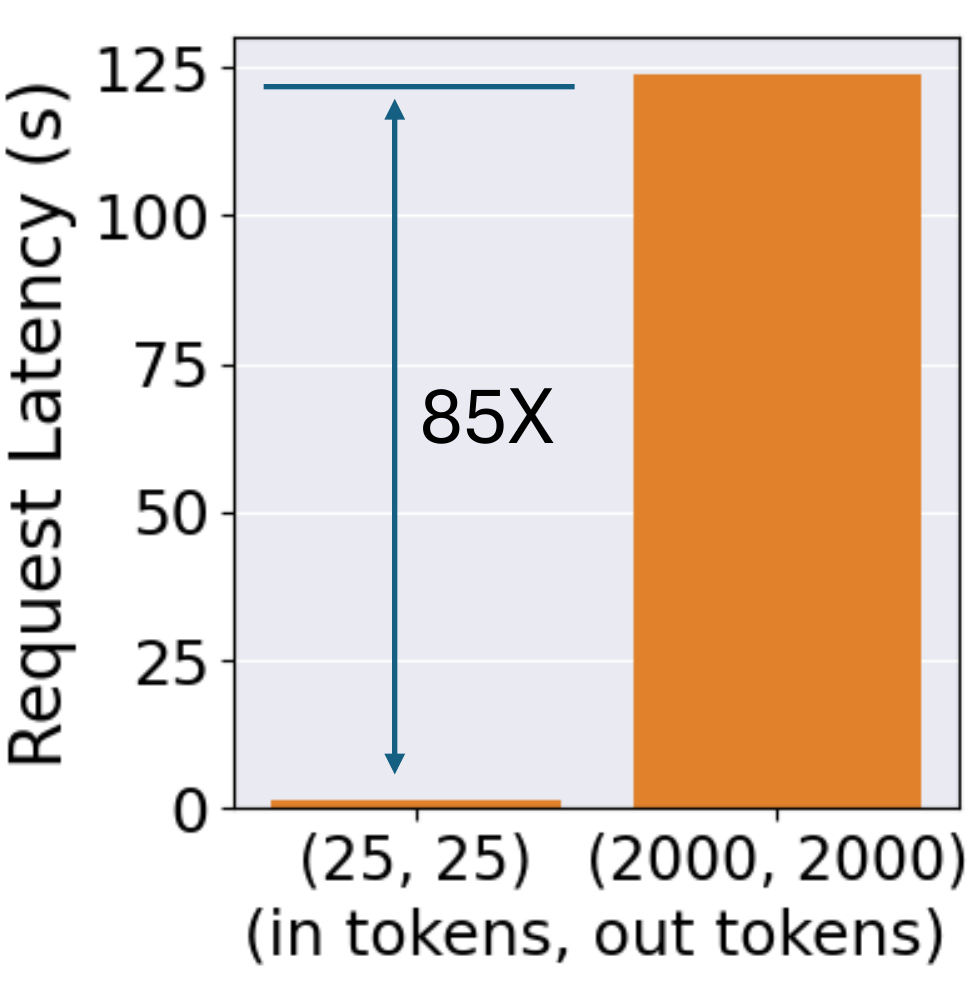}
    \caption{LLaMA-70B}
    \end{subfigure}
    \caption{Request latency of different input/output lengths on A100-80G.}
    \label{fig:llm-diff-workload}
\end{figure}

Unlike traditional machine learning workloads, LLM tasks exhibit significant variance in \textit{request sizes}, defined by input and output lengths. For example, ResNet~\citep{he2015deep} requires a fixed-dimension input (image size) and generates a fixed-dimension output (classification size). Conversely, transformer-based language models are flexible to support variable-length prompts and produce variable-length generation sequences. For instance, Figure \ref{fig:dataset} illustrates the request size distributions of Chatbot Arena, demonstrating the extensive diversity of request sizes in practical scenarios. 
As a result, high variance in request sizes introduces significant variation in request latency. As illustrated in Figure~\ref{fig:llm-diff-workload}, request latency can increase by $110\times$ when the input/output length expands from 25 tokens to 2000 tokens for the Llama2-7B model served on an A100 GPU. Consequently, it is crucial to recognize that LLM requests, unlike non-autoregressive models, impose varied loads on GPU resources.

\section{GPU Cost Efficiency Analysis}
\label{sec:observation}
In this section, we analyze GPU cost efficiency for LLM services by serving Llama2-7b on NVIDIA A100~\citep{a100-spec} and A10G~\citep{a10-spec} as a representative example. We show that GPU cost efficiency is influenced by three key LLM service characteristics: request size (\S~\ref{sec:ob-req-size}), latency SLO (\S~\ref{sec:ob-slo}), and request rate (\S~\ref{sec:ob-request-rate}). 
For each characteristic, we demonstrate opportunities to exploit the heterogeneity of GPU types to increase cost efficiency and reduce deployment cost.
Each plot is tagged with the request size, request rate, and SLO used to generate the plot.
We use vLLM-0.2.7 as the serving engine~\citep{kwon2023efficient}. 

\subsection{Definitions}
\label{sec:observation-definitions}
\textbf{Service-level Objective (SLO).} 
SLOs are performance targets that define the acceptable quality of service, and a specific SLO varies according to the service’s interactivity needs.
As in prior work~\citep{kwon2023efficient, zhong2024distserve, yu2022orca}, we use the average \textit{Time Per Output Token (TPOT)} as our SLO. TPOT is determined by dividing request latency by the number of generated tokens. SLOs are application dependent: in-line code editors (e.g., GitHub Copilot~\citep{copilot}) require tight latency deadlines to suggest real-time code additions, whereas summarization services may permit additional processing time.
There are other common definitions of SLO, such as time to first token and request latency, and Mélange is flexible to support these and other alternative definitions of SLO.

\noindent \textbf{Cost Efficiency Metric.} 
We use \textit{tokens per dollar} (\tperd) to measure GPU cost efficiency, calculated by summing input and output tokens and dividing the total by the GPU's on-demand rental cost for a given time period. Cost models are orthogonal to Mélange; we chose this cost model for its simplicity, but cost efficiency can be computed with alternative formulations without affecting Mélange's efficacy.
In general, we derive \tperd by finding the input and output token rates while at the highest GPU saturation for which TPOT still meets a specified SLO.

\subsection{Request Size and Cost Efficiency}
\label{sec:ob-req-size}

\begin{figure} 
    \centering
    \begin{subfigure}[t]{0.44\linewidth}
        \includegraphics[width=\linewidth]{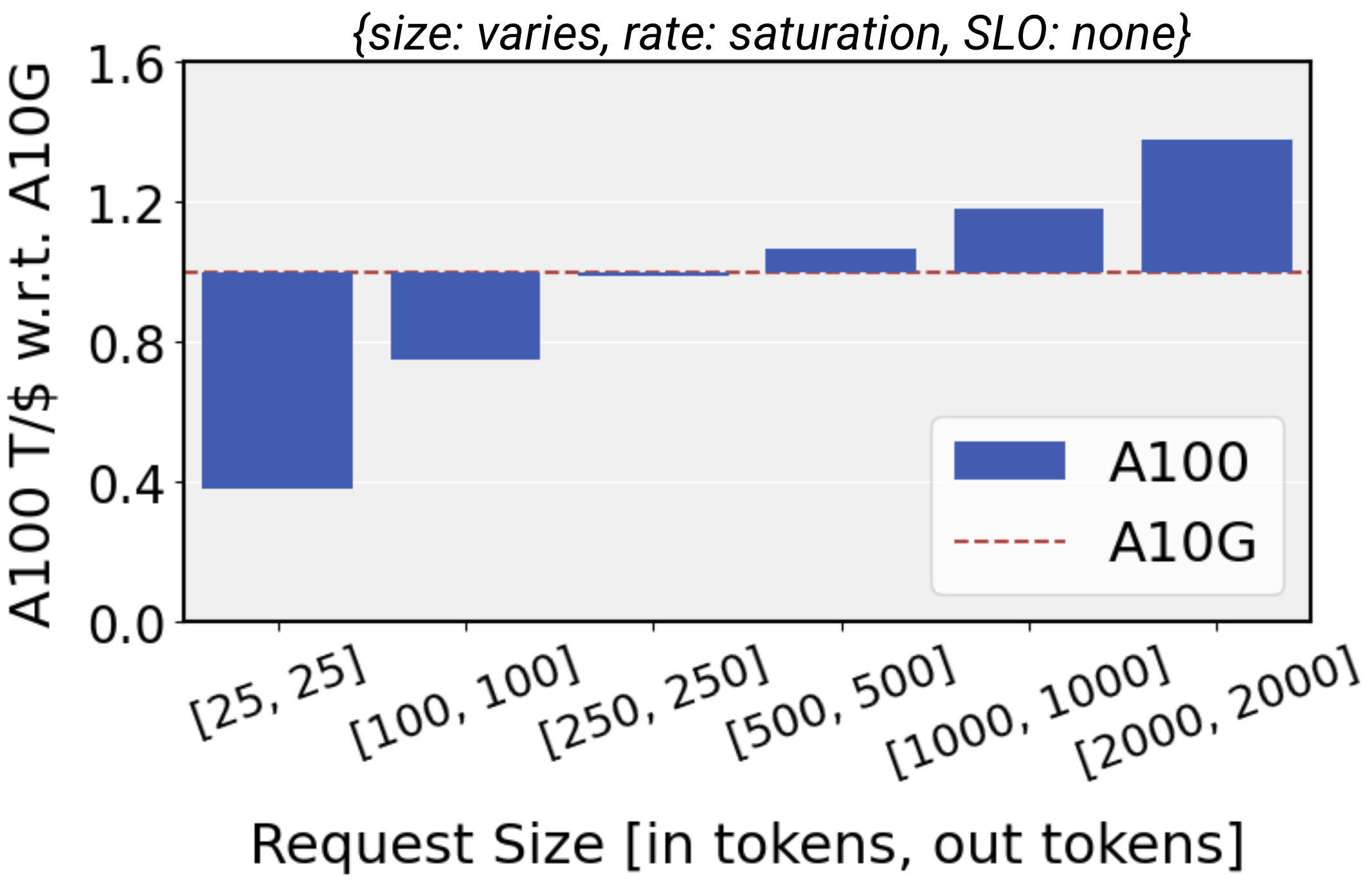}
        \caption{Equivalent input and output lengths}
        \label{fig:dollar-tput-a10-a100}
    \end{subfigure}
    \hspace{4pt}
    \begin{subfigure}[t]{0.44\linewidth}
        \includegraphics[width=\linewidth]{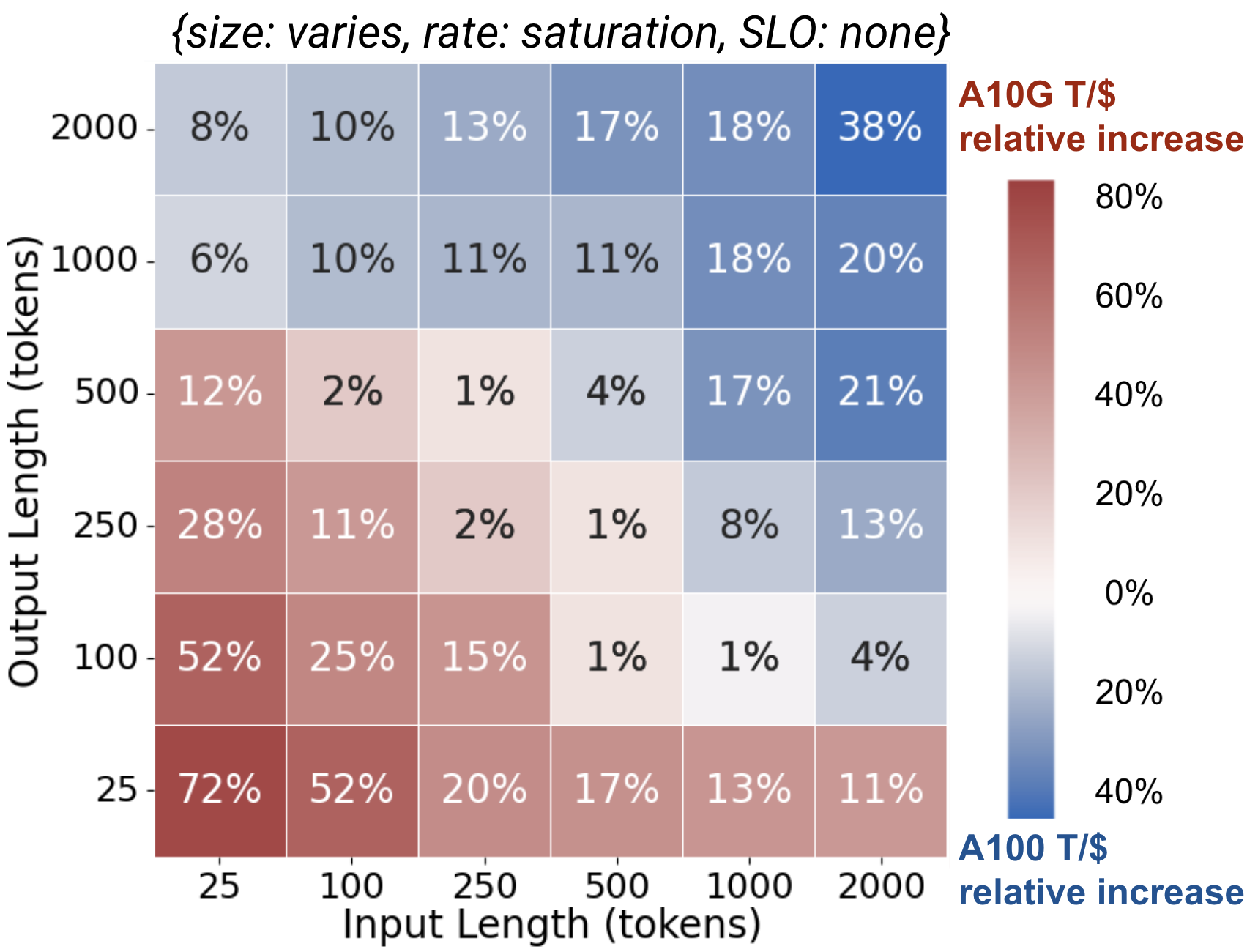}
        \caption{Input and output lengths vary independently}
        \label{fig:workload-a10-a100}
    \end{subfigure}
    \caption{Figure (a) depicts A10G and A100's relative \tperd across request sizes. Figure (b) expands (a) into separate input and output length dimensions. Tile colors indicate which GPU achieves higher \tperd, and values represent the percent increase of \tperd relative to the less cost efficient GPU.}
    \label{fig:a10-a100-tputs}
\end{figure}

Unlike many traditional DNNs, LLMs exhibit significant variance in model request sizes (input and output lengths)~\citep{patel2023splitwise}. In this section, we show that request size variance influences GPU cost efficiency and can even determine which GPU is most cost efficient.

\textbf{Experiment:} We serve Llama2-7b on A10G and A100 GPUs, and derive each GPU's \tperd at maximum GPU saturation across a range of request sizes (~\autoref{fig:dollar-tput-a10-a100}). Interestingly, no single GPU consistently delivers the highest tokens per dollar (\tperd) across all request sizes. Instead, both GPUs are most cost efficient in separate regions of the request size spectrum. For smaller request sizes, A10G exhibits up to $2.6\times$ greater \tperd than A100. Conversely, for larger request sizes, A100 achieves up to $1.5\times$ the cost efficiency of A10G.

We extend this exploration to show the separate impacts of input and output lengths on \tperd (~\autoref{fig:workload-a10-a100}). Each dimension influences cost efficiency similarly: smaller sizes are best served on A10G, and larger sizes are best served on A100. 
Note that the difference can be significant, as using a single GPU type to serve requests across the entire request size space misses opportunities to produce up to 72\% more output tokens for the same cost. This reveals the opportunity to use a mix of GPU types to serve requests for which they are most cost effective.

\begin{figure} 
    \centering
    \begin{subfigure}[t]{0.44\linewidth}
        \includegraphics[width=\linewidth]{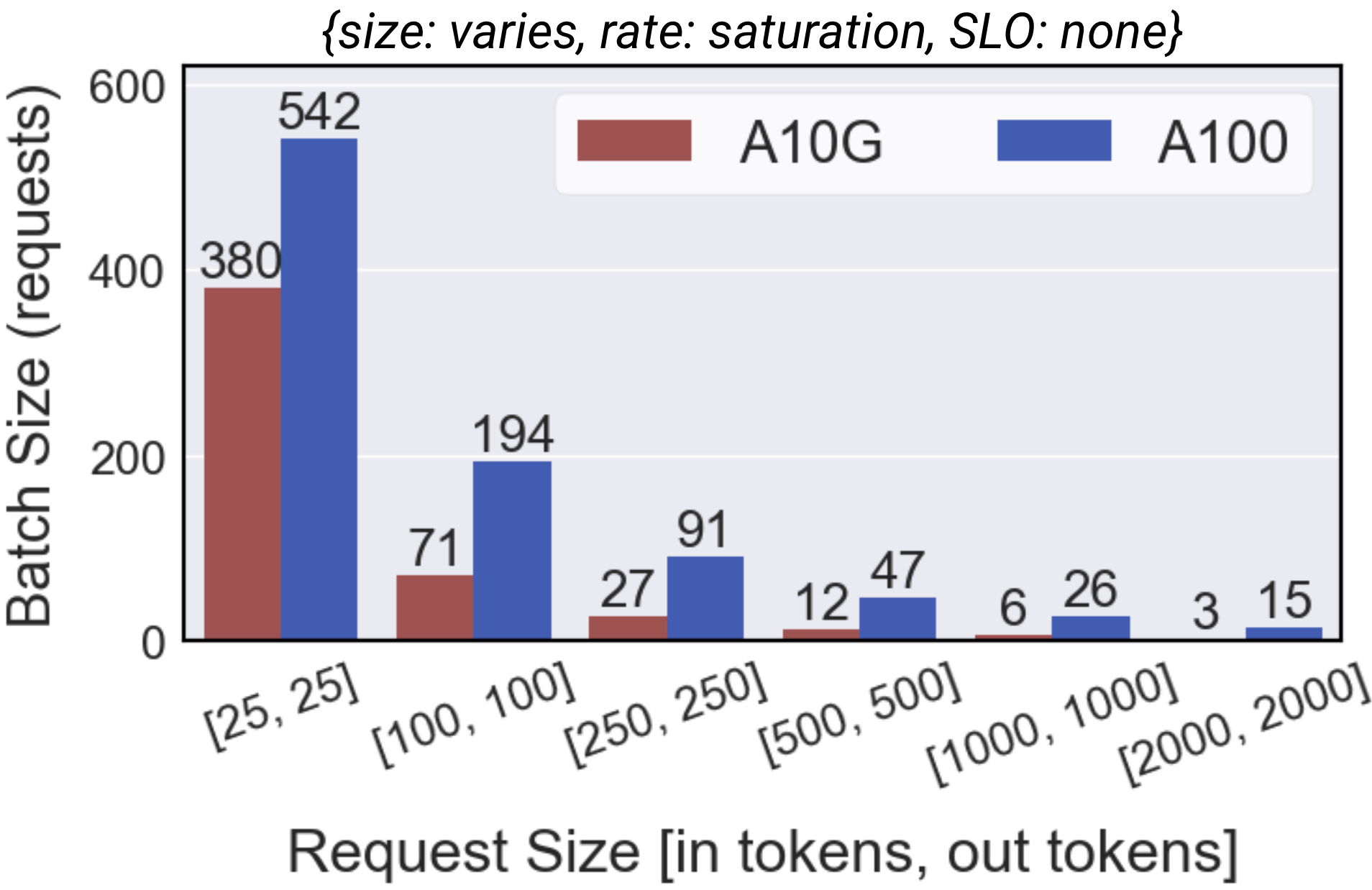}
        \caption{Absolute batch sizes}
        \label{fig:absolute-batch-sizes}
    \end{subfigure}
    \hspace{9pt}
    \begin{subfigure}[t]{0.44\linewidth}
        \includegraphics[width=\linewidth]{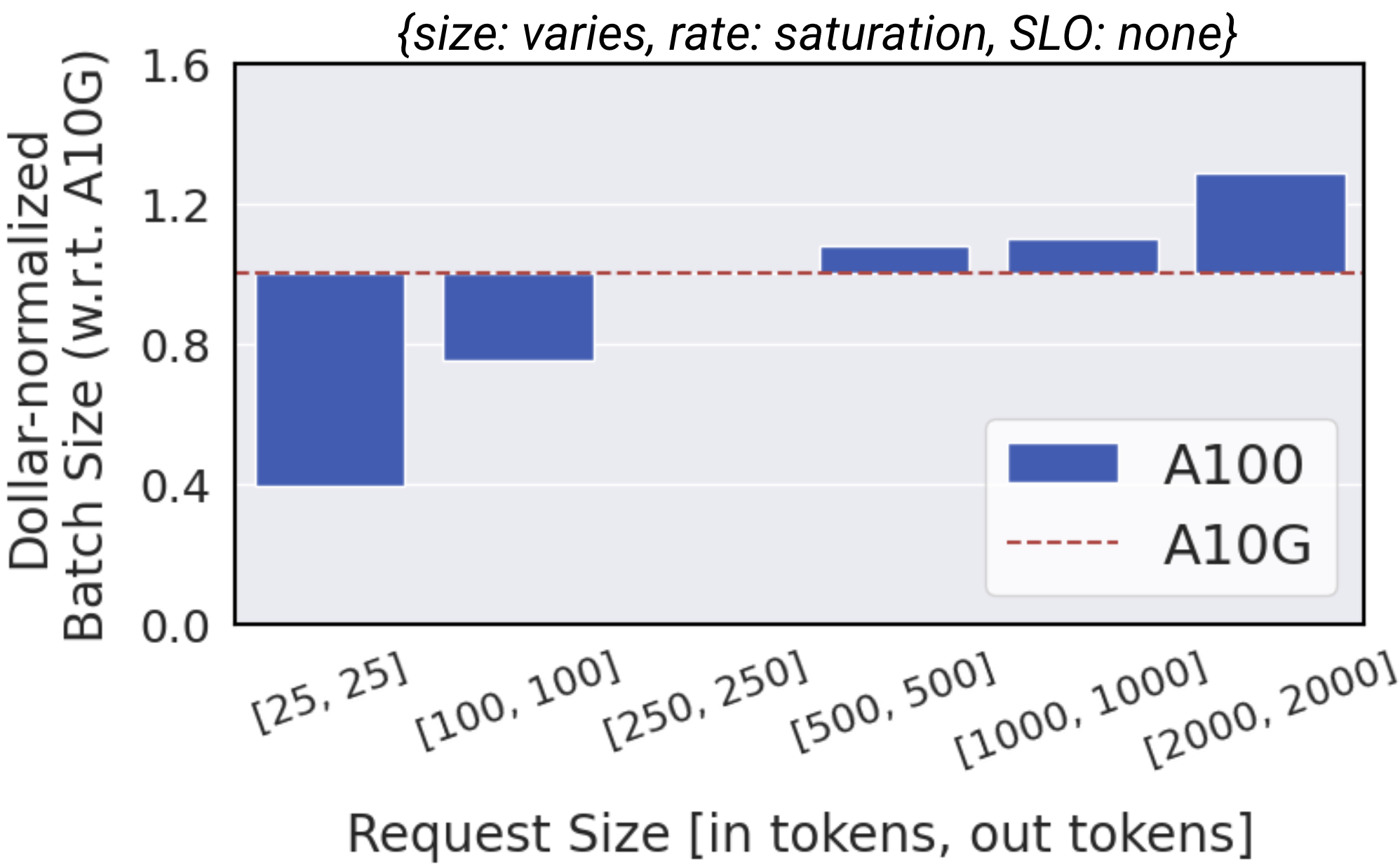}
        \caption{Dollar-normalized batch sizes}
        \label{fig:normalized-batch-sizes}
    \end{subfigure}
    \caption{(a) depicts the absolute batch sizes of A10G and A100 serving Llama2-7b at maximum saturation, (b) reports the same batch sizes divided by GPU cost, plotting with respect to A10G.}
    \label{fig:all-batch-sizes}
    % \vspace{-1.2em}
\end{figure}

\textbf{Source of Cost Efficiency Gains:} 
To isolate how request size influences relative cost efficiency,
we examine request size's effects on batch size, which serves as a proxy for throughput. ~\autoref{fig:all-batch-sizes} depicts absolute batch sizes and batch sizes normalized by instance cost of each GPU at maximum saturation. 

A10G and A100 have similar cost-normalized batch sizes at 250 input/output tokens, but as the request size increases to 2K input/output tokens, A10G's absolute batch size decreases by $9\times$ whereas A100's only decreases by $6\times$ due to its superior memory size and bandwidth. As a result, A100's cost efficiency advantage over A10G increases with the increase in request size.  
In contrast, reducing the size from 250 to 25 input/output tokens expands A10G's batch size by $15.2\times$, whereas A100's growth is $5.89\times$.Because A100's batch sizes are larger, A100 is more significantly constrained by per-request latency overheads (e.g., due to interference of prefill and decode~\citep{hu2024inference}) 
As a result, A10G's cost-normalized batch size exceeds A100's at short request lengths, leading to greater overall \tperd. 

\begin{figure} 
    \centering
    \begin{subfigure}[t]{0.44\linewidth}
        \includegraphics[width=\linewidth]{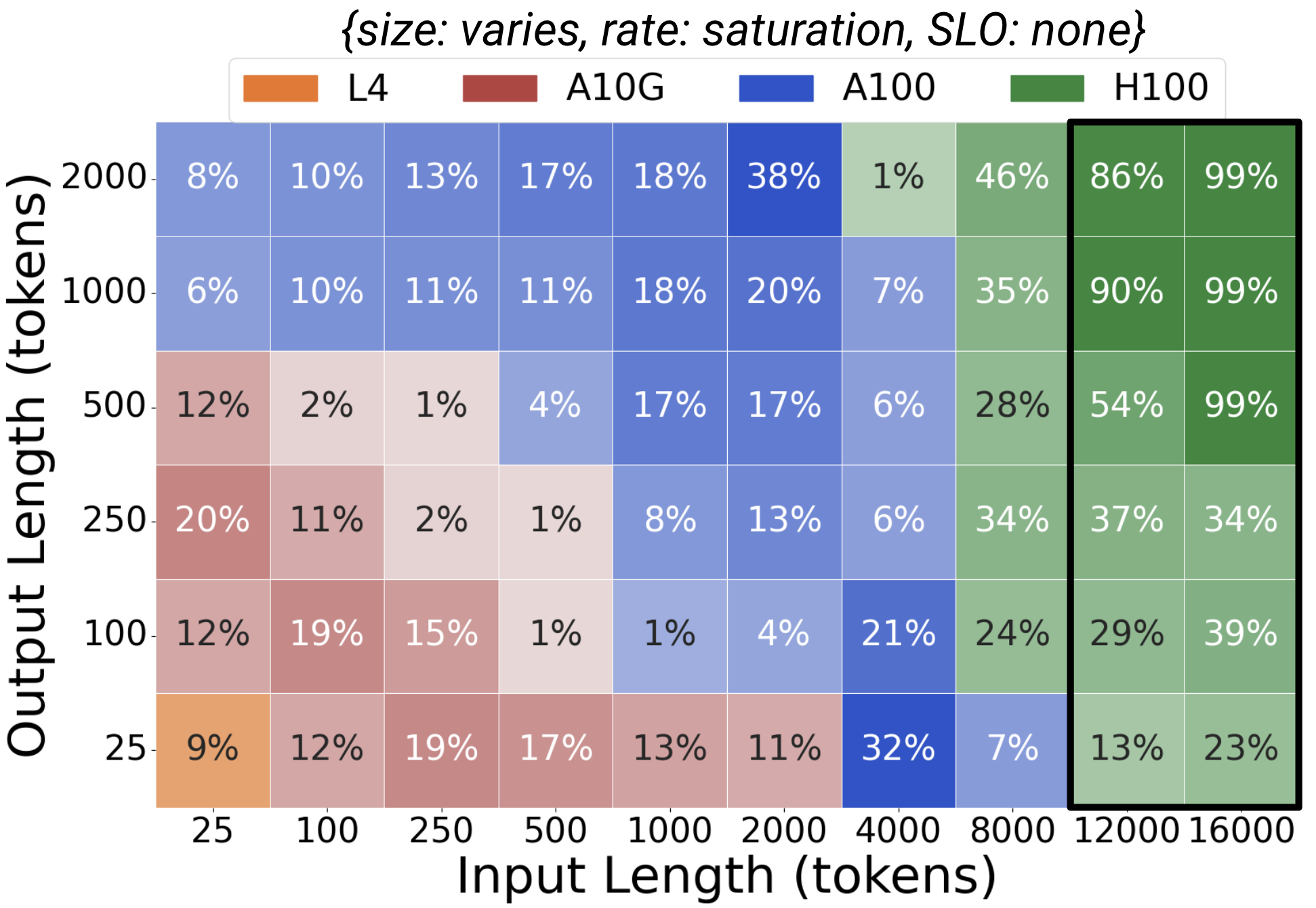}
        \caption{Best GPU relative to second best GPU}
        \label{fig:all-gpus-best}
    \end{subfigure}
    \begin{subfigure}[t]{0.44\linewidth}
        \includegraphics[width=\linewidth]{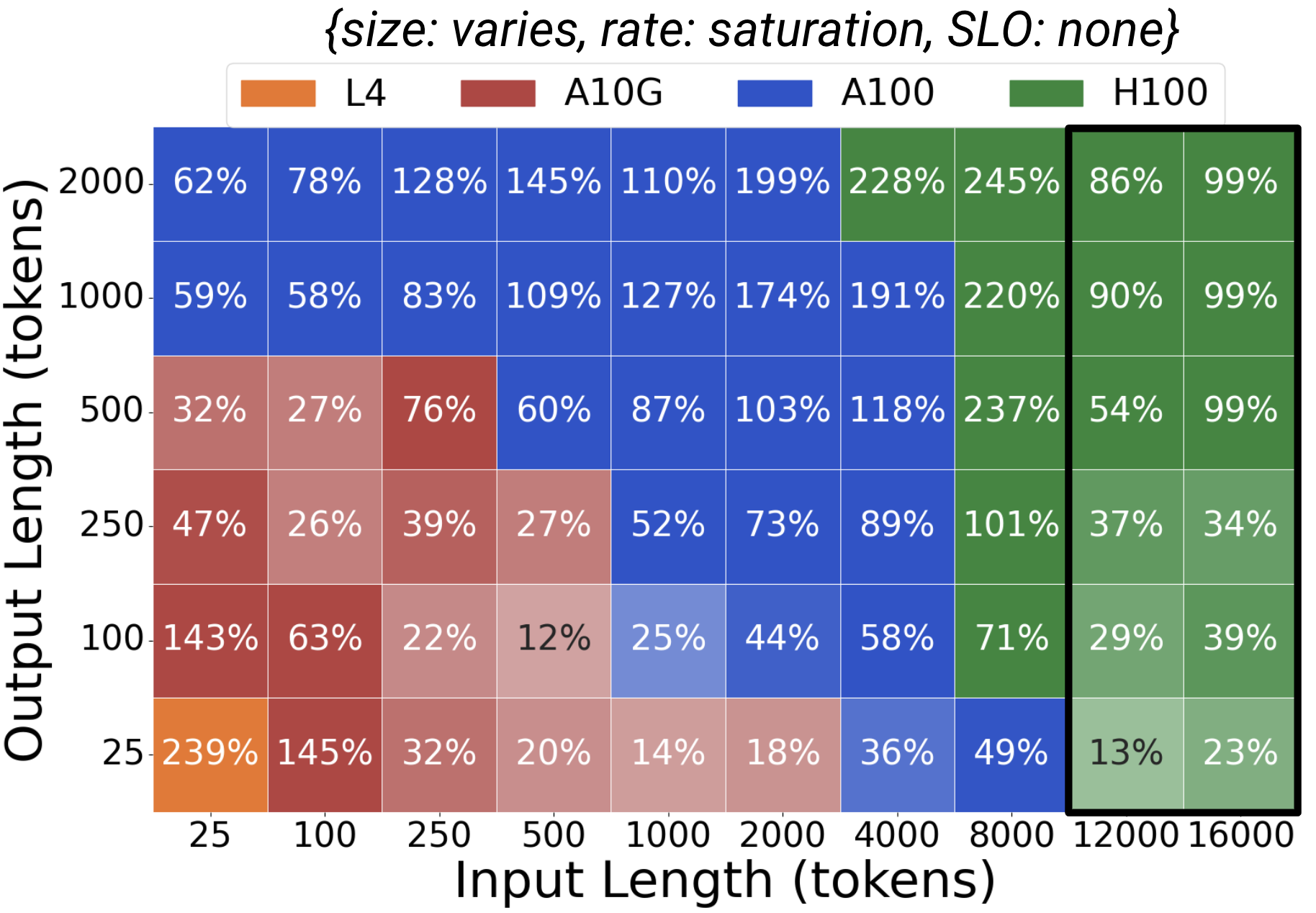}
        \caption{Best GPU relative to worst GPU}
        \label{fig:all-gpus-worst}
    \end{subfigure}
    \caption{Comparison of L4, A10G, A100, and H100. Tile colors indicates the GPU with greatest \tperd. (a) tile values are the \tperd  \%-increase of the best GPU compared to the second best for that tile. (b) compares the best GPU to the worst GPU. In black boxes, only A100 and H100 are compared.}.
    \label{fig:all-gpus}
    \vspace{-2.2em}
\end{figure}

\begin{figure}[ht]  
    \centering
    \begin{minipage}{0.48\textwidth}
        \includegraphics[width=\linewidth]{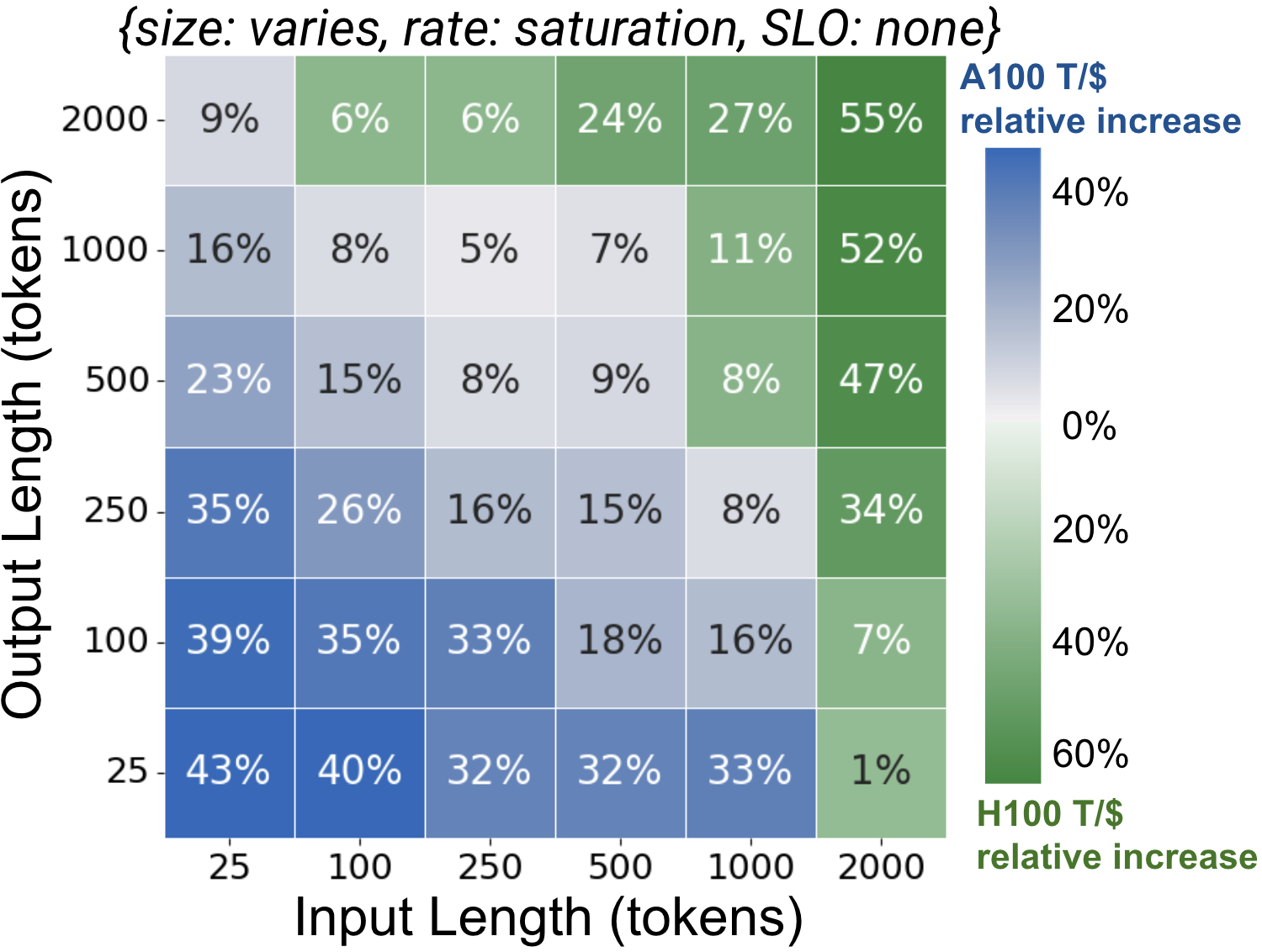}
        \caption{\tperd comparison between H100x2 and A100x2 serving Llama2-70b. }
        \label{fig:h100-v-a100}
    \end{minipage}
    \hspace{8pt}
    \begin{minipage}{0.45\textwidth}
        \includegraphics[width=\linewidth]{figures/req-rate-comparison.png}
        \caption{GPU on-demand cost for three GPU provisioning strategies.}
         \label{fig:request-rate}
    \end{minipage}
    \vspace{-1.5em}
\end{figure}

\textbf{Other Hardware and Model Size}
We extend our analysis to more GPU types and a larger model variant (Llama2-70b). ~\autoref{fig:all-gpus} depicts the relative cost efficiency across four GPU types. Once again, as request sizes increase, we observe a progression of the most cost efficient GPU from lower-end to higher-end GPUs, matching our observations above. 
Similar trends are observed in the larger Llama2-70B model when comparing H100 and A100 GPUs, as detailed in ~\autoref{fig:h100-v-a100}.

\noindent \textbf{Key Takeaways:} There is no universally most cost-efficient GPU for a given LLM. Instead, GPU cost efficiency is highly dependent on request sizes. 
Lower-end GPUs are more cost-effective for small request sizes whereas higher-end GPUs are best for large request sizes.
These findings generalize to settings with more GPU types and larger model sizes.

\subsection{SLO and Cost Efficiency}
\label{sec:ob-slo}

\begin{figure}[ht]  
    \centering
    \begin{minipage}{0.43\textwidth}
        \includegraphics[width=\linewidth]{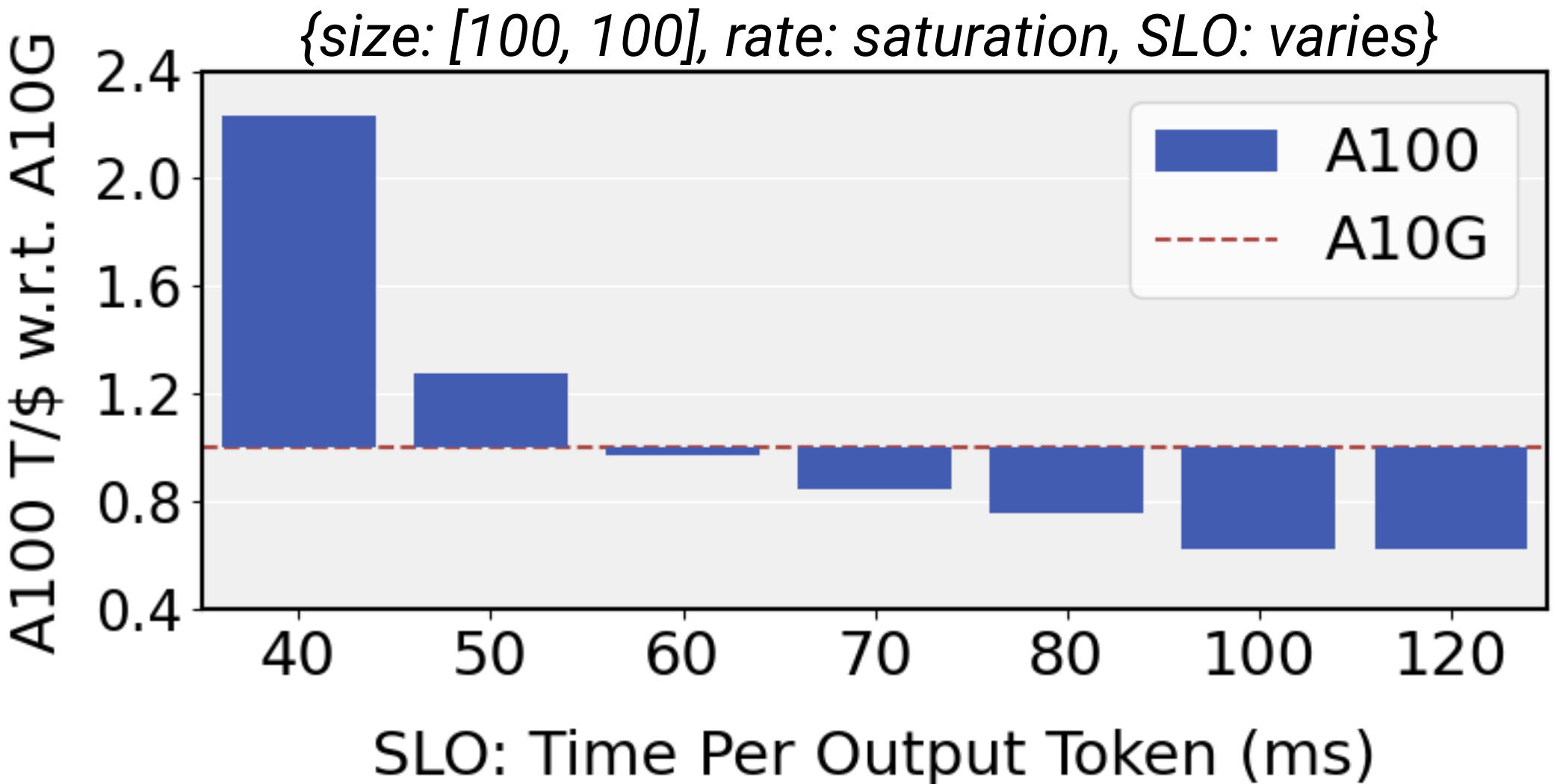}
        \caption{\tperd comparison between A10G and A100 across a range of TPOT SLO parameters.}
        \label{fig:slo}
    \end{minipage}
    \hspace{8pt}
    \begin{minipage}{0.50\textwidth}
        \includegraphics[width=\linewidth]{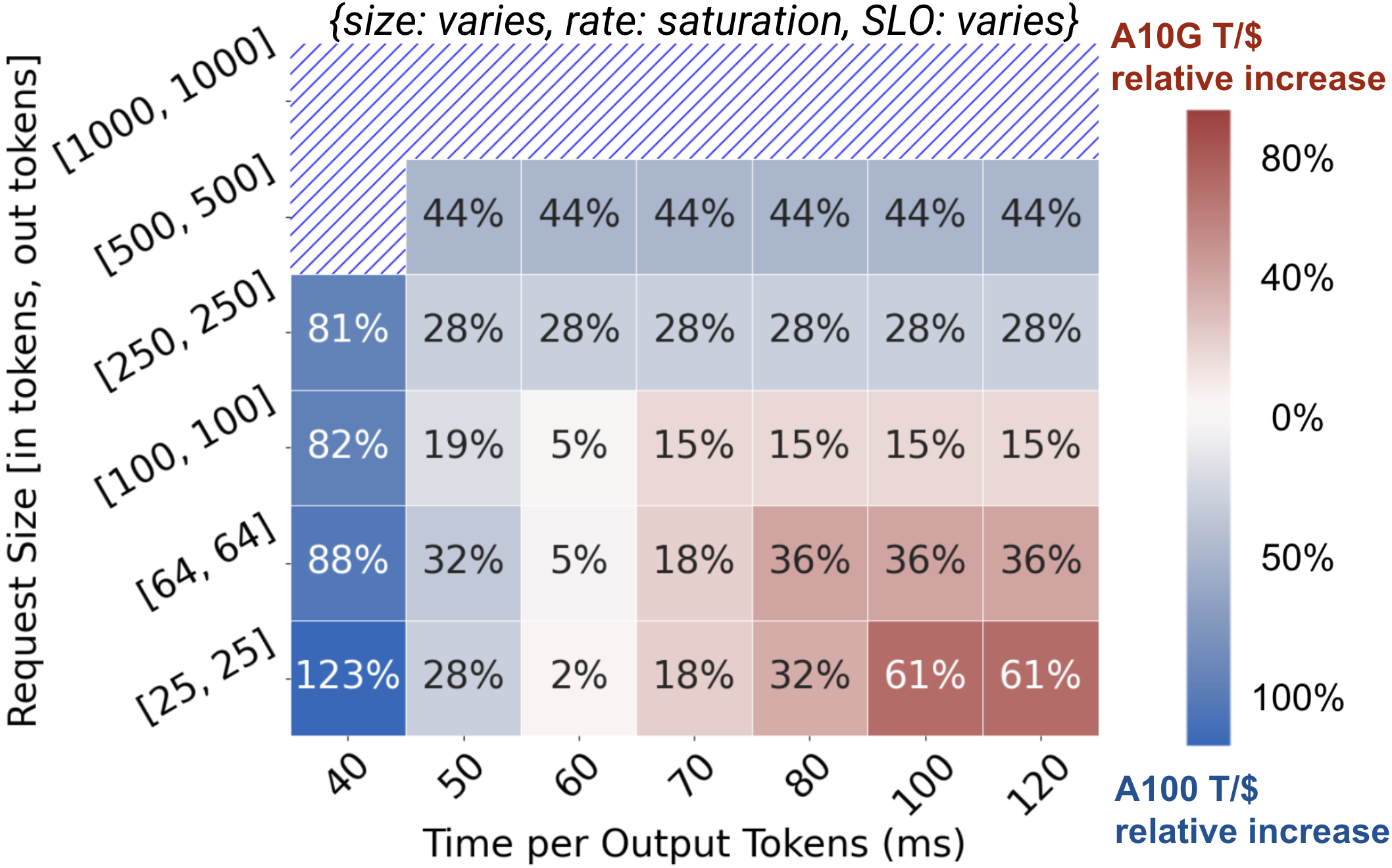}
        \caption{Relative increase in \tperd when combining SLO and request size.} 
        \label{fig:qps-slo-2}
    \end{minipage}
    % \vspace{-2em}
\end{figure}

In this section, we examine the impact of TPOT SLOs on GPU cost efficiency and highlight the joint effects of SLO and request size.

\textbf{Experiment:} We serve Llama2-7b on A10G and A100 and measure \tperd by maximally saturating each GPU while keeping TPOT below SLO, repeating this across several TPOT deadlines (~\autoref{fig:slo}).
Under tight SLO constraints ($<$60ms), A100 demonstrates significantly greater \tperd than A10G ($2\times$). 
A10G's higher processing latency restricts the throughput it can achieve within a tight TPOT deadline, while A100 maintains much higher throughput even at low latency. However, as the SLO gradually loosens (60-160ms), A10G's higher latency is less problematic, dramatically increasing its \tperd and surpassing that of A100 (by $>40\%$). Importantly, this example uses a small request size (64 input/output tokens), which was shown in \S~\ref{sec:ob-req-size} to be best served on A10G. However, a tight SLO degrades A10G's cost efficiency much more severely than A100's and pushes the advantage to A100, exemplifying the tight interplay between SLO and request size explored further below.

\textbf{SLO and Request Size Interplay:} ~\autoref{fig:qps-slo-2} presents relative cost efficiency between A10G and A100 for a broad range of TPOT SLOs and request sizes. 
At tight SLOs (40-60ms), A100 always has higher \tperd (up to $2\times$). At 80ms, A10G begins showing modest benefit over A100 for small request sizes. Finally, at 100-160ms, A10G demonstrates much greater \tperd advantage over A100 for the same request sizes (up to $1.5\times$), yet A100 is always more cost efficient for larger requests.
As demonstrated, a modification to TPOT SLO shifts the boundary within the request size space between which different GPU types are most cost effective and significantly influences the magnitude of cost efficiency differences between GPUs. As a result, both request size and SLO must be considered in tandem when determining cost efficiency.

\noindent \textbf{Key Takeaways:} To meet strict SLOs, expensive GPUs are necessary due to the higher latency of cheaper GPUs. However, as SLO is loosened, lower-end GPUs can be used to cut deployment costs. 

\subsection{Request Rate and Cost Efficiency}
\label{sec:ob-request-rate}
In this section, we investigate the relationship between request rate and GPU cost efficiency.

\textbf{Experiment:} ~\autoref{fig:request-rate} illustrates the cost of serving Llama2-7b at a range of request rates using three strategies: A10G-only, A100-only, or a mix of both. The y-axis is absolute cost instead of \tperd because each provisioning strategy serves the same request rates and thus the same number of tokens; only the cost varies across strategies. 

As request rate increases, A100-only is increasingly more cost effective than A10G-only. This is because the requests are of size [1000 in tokens, 250 out tokens], which \S~\ref{sec:ob-req-size} shows is more cost effective on A100. However, A10G-only still presents benefits at low request rates ($0$-$1$ req/s). Periods of idleness or low activity are common in real-world services~\citep{romero2021infaas}, and the service should right-size to a cheaper GPU (here, A10G) when a higher-end GPU (here, A100) is drastically underutilized.  

\textbf{Mixing GPU Types:} The hybrid approach of serving the model on both A10G and A100 GPUs consistently yields the lowest deployment cost. Because A100s have such large capacity, scaling with only A100s is coarse-grained and often leads to underutilized resources. Instead, A10Gs and A100s can be mixed such that A100s satisfy the bulk of the service demands, while A10Gs handle the remaining load at reduced cost. ~\autoref{fig:request-rate} highlights a case where using 2 A100s and 1 A10G results in a 24\% cost saving over A100-only and 31\% over A10G-only.

\textbf{Key Takeaways:} During low activity periods, LLM services should right-size to cheaper low-end GPUs. Provisioning a mix of GPU types enables finer-grained resource scaling, which better aligns the allocated GPU capacity with request load. This increases GPU utilization and consistently achieves lowest serving cost.

\section{Mélange: Automating Cost-Efficient GPU Selection}
\label{sec:algorithm}
Building on the observations in \S~\ref{sec:observation} that request size, request rate, and SLO all jointly determine GPU cost efficiency, we present Mélange, an allocation framework that considers each of these three dimensions in-tandem to derive the minimal-cost GPU allocation that meets an LLM service's request load while adhering to SLO constraints.
~\autoref{fig:melange-diagram} depicts the Mélange framework. Mélange flexibly supports any GPU type (\greycircle{1a}) and LLM service definition (\greycircle{1b}), uses a one-time offline profiling step to measure GPU performance (\greycircle{2}), formulates the task of GPU allocation as a bin packing problem (\greycircle{3}), then computes the minimal-cost GPU allocation (\greycircle{4}).

\subsection{Problem Formulation}
\label{sec:problem-formulation}

We begin by defining the key terms utilized in our problem formulation and solution. An LLM service \textbf{workload} is characterized by its overall request rate along with a distribution of input and output sizes. A distribution of request sizes is used rather than fixed values due to the inherent variability of LLM request sizes. Specifically, 
a workload is a histogram where each bucket corresponds to a range of request sizes and a bucket's value is the request rate of requests within the bucket's size range. The service \textbf{cost} is computed by summing the hourly on-demand cloud renatl rates for each of the selected GPUs. We define \textbf{SLO} based on average TPOT, however, Mélange can be extended to other definitions of SLO such as time to first token (TTFT).

\textbf{Problem Definition:} Given a workload, GPU costs, and SLO requirements, our objective is to provision GPUs that can minimize deployment cost while adhering to latency SLO constraints. 

\subsection{Inputs} Mélange takes as input the set of available GPU types (\greycircle{1a}) and the LLM service definition (\greycircle{1b}) made up of the workload profile and SLO. Each of these inputs can be modified, such as adding a new hardware accelerator or redefining SLO based on end-to-end request latency, and Mélange's downstream components still derive the minimal-cost allocation. Due to the large diversity of hardware accelerators and LLM services, Mélange's extensibility is critical for usability.

\subsection{Offline Profiling}
A one-time offline profiling step (\greycircle{2}) is required to measure the performance of each GPU. For each request size bucket in the workload histogram, we gradually increase the request rate until the GPU is saturated. We record per-request TTFT and TPOT as the request rate is increased, which are sufficient metrics to capture the timing behavior of a request end-to-end~\citep{liu2024andes}. 
Then, given an SLO, Mélange can quickly find the maximum throughput each GPU achieves across request sizes while adhering to the SLO. Empirically, the one-time profiling is not time-consuming (<1hr).

\subsection{Allocation Algorithm}
The allocation algorithm's (\greycircle{3}) objective is to map the workload to a minimal-cost set of GPUs that are constrained by adhering to SLO. Our insight is that this task can be formulated as a cost-aware variant of the bin packing problem. Mélange partitions workload buckets into smaller \textit{slices} for fine-grained packing, then assigns the slices (items) to GPUs (bins). We first define a slice (\S~\ref{sec:bucket-slice}), compute the load of a slice (\S~\ref{sec:percent-load}), then create the ILP formulation (\S~\ref{sec:ilp}). 

\subsubsection{Request Buckets and Slices}
\label{sec:bucket-slice}
A workload histogram has two dimensions, input length and output length, and each histogram bucket's value is the aggregate request rate for requests within the bucket's size range.
We further break each bucket down into slices for finer-grained bin packing. 
A parameter, \textit{slice factor}, indicates the number of slices that each bucket is divided into. In a setting with a slice factor of 8 and a bucket with a request rate of 4, the bucket would be segmented into 8 slices each corresponding to a request rate of 0.5 requests/s.
The slice factor can be tuned to reach the desired balance between granularity and solution complexity, but we have not found overall performance to be sensitive to slice factor.

\subsubsection{Load}
\label{sec:percent-load}
The solver requires an estimate of the load of each slice to ensure that a GPU's capacity is not exceeded and SLO is not violated. The load of a slice with request size $s$ and rate $r$ on GPU $G$ is calculated as \(\frac{r}{MaxTput(G, s, SLO)}\), where $MaxTput(G, s, SLO)$ is the maximum request/s $G$ can achieve for requests of size $s$ while adhering to $SLO$. For instance, if \(MaxTput(G, s, SLO)=10\, reqs/s\) and $r = 1$, the load is calculated as \(1 / 10 = 0.1\) . Each GPU's maximum capacity is defined as 1. This approximation allows us to calculate the aggregate load of slices with differing sizes and rates. 
Based on offline profiling, we compute $MaxTput(G, s, SLO)$ for each bucket in the workload histogram. 

\subsubsection{ILP Formulation}
\label{sec:ilp}
We formulate the ILP with two decision variables. First, let \(A\) be a matrix $\{0, 1\}^{N \times M}$, where \(N\) is the number of slices, and \(M\) is the number of GPU types. \(A_{i,j}=1\) if slice \(i\) is assigned to GPU type \(j\), and \(0\) otherwise. 
The second decision variable, \(B\), is a vector $\mathbb{Z}_{\geq 0}^M$ of non-negative integers, where \(B_j\) specifies the number of GPUs of type \(j\) to be allocated. 
$L$ is a matrix of size ${N \times M}$ where $L_{i,j} \in [0, 1]$ is the fractional load of slice $i$ on GPU type $j$. $L$ is computed offline by the process described in \S~\ref{sec:percent-load}.
\(c_j\) denotes the cost of GPU type \(j\).

\noindent
\begin{minipage}[t]{0.45\textwidth} % [t] for top alignment

Our objective is to minimize the total GPU allocation cost:

\vspace{8pt}

The ILP constraints are as follows. First, each task slice is assigned to exactly one GPU type:

\vspace{8pt}

Second, for each GPU type, the number of GPUs designated in vector $B$ must satisfy the cumulative load prescribed to it in matrix $A$:

\vspace{8pt}

Lastly, elements of matrix $A$ are binary, and elements of vector $B$ are non-negative:
\end{minipage}%
\hfill
\begin{minipage}[t]{0.55\textwidth} % [t] for top alignment
    \small % Adjust the font size (e.g., \small, \footnotesize, \scriptsize, \tiny)
    \begin{equation}
        \arg\min_{B}(\sum_{j=1}^M B_j \cdot c_j)
    \end{equation}
    \vspace{-7pt} % Adjust vertical space between equations
    \begin{equation}
        \forall i \in \{1, \ldots, N\}, \quad \sum_{j=1}^M A_{i,j} = 1
    \end{equation}
    \vspace{-7pt} % Adjust vertical space between equations
    \begin{equation}
        \forall j \in \{1, \ldots, M\}, \quad \sum_{i=1}^N A_{i,j} \cdot L_{i,j} \leq B_j
    \end{equation}
    \vspace{-7pt} % Adjust vertical space between equations
    \begin{equation}
        \forall i, \forall j, \quad A_{i,j} \in \{0, 1\}
    \end{equation}
    \vspace{-7pt} % Adjust vertical space between equations
    \begin{equation}
        \forall j \in \{1, \ldots, M\}, \quad B_j \geq 0
    \end{equation}
\end{minipage}

The solution is computed using an off-the-shelf solver~\citep{pulp2023}. Upon solution, the decision variable $B$ holds the minimal-cost GPU allocation (\greycircle{4}) that meets the workload demand and adheres to SLO. 
\section{Evaluation}

\label{sec:experiment}
We assess Mélange's performance across diverse hardware, request sizes, rates, and SLOs. Mélange consistently achieves significant cost savings (up to 77\%) compared to single-GPU-type strategies, and the selected allocations successfully attain TPOT SLO for over 99.5\% of requests.

\subsection{Experiment Setup}
\textbf{Environment.}
\label{sec:exp-setup}
We use four NVIDIA GPU types that capture a broad range of prices and specifications, with details in ~\autoref{tab:hardware}. In increasing price order, we use L4, A10G, A100-80G, and H100. 
To determine the GPU cost, we select the lowest on-demand price available from major cloud providers (AWS, Azure, and GCP). Since on-demand H100 is not offered by these major providers, we defer to the pricing from RunPod~\citep{runpod} due to its popularity and availability. To ensure fair cost comparisons, we normalize RunPod's H100 pricing to match the pricing structures of major platforms. We calculate this by comparing RunPod's H100 cost (\$4.69) to RunPod's A100-80G cost (\$2.29), then adjusting relative to the A100's price on major clouds (\$3.67), resulting in a normalized price of $(4.69 / 2.29) \times 3.67 = \$7.516$ for H100. In each experiment, we serve Llama2-7b~\citep{touvron2023llama2} with vLLM 0.2.7~\citep{kwon2023efficient}. 

\begin{table}[h]
    \centering
    \begin{small}
    \begin{tabular}{c|c|c|c|c}
    \toprule
    Type                  & L4   & A10G (PCIe) & A100-80G (SXM) & H100 (SXM) \\
    \midrule
    On-demand Price (\$/h)          & 0.7  & 1.01        & 3.67           & 7.516\footnotemark        \\
    % Normalized Cost       & 0.693 & 1 & 3.6634 & 7.5028 \\
    Instance Provider     & GCP  & AWS         & Azure          & RunPod     \\
    Instance Name         & g2-standard-4 & g5.xlarge & NC24ads\_A100\_v4/N.A.    & N.A.        \\
    Memory (GB)           & 24            & 24        & 80      & 80         \\
    Memory Bandwidth (GB/s)  & 300        & 600       & 1935    & 3350       \\
    FP16 (TFLOPS)         & 242           & 125       & 312     & 1979       \\
    \bottomrule
    \end{tabular}
    \caption{Specifications of four NVIDIA GPUs: L4, A10G, A100, and H100.}
    \label{tab:hardware}
    \end{small}
\end{table}

\begin{figure}[!htbp] 
    \centering
    \begin{subfigure}[b]{0.40\linewidth}
        \includegraphics[width=\linewidth]{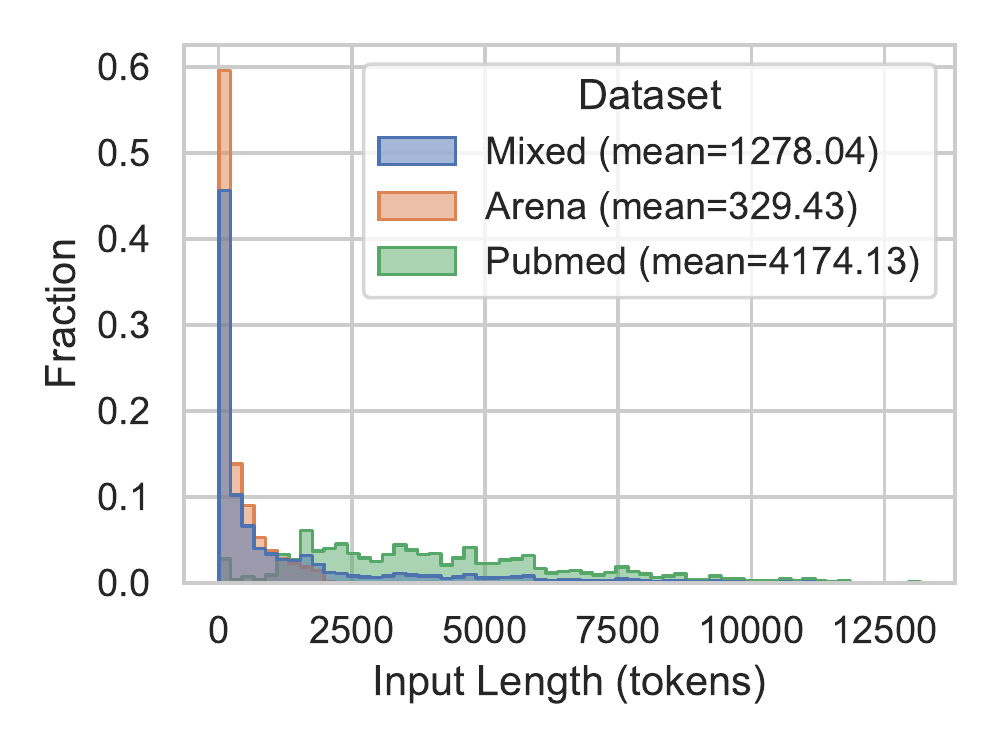}
        \captionsetup{skip=-4pt}
        \caption{Input length distributions.}
        \label{fig:data-input}
    \end{subfigure}
    \begin{subfigure}[b]{0.40\linewidth}
        \includegraphics[width=\linewidth]{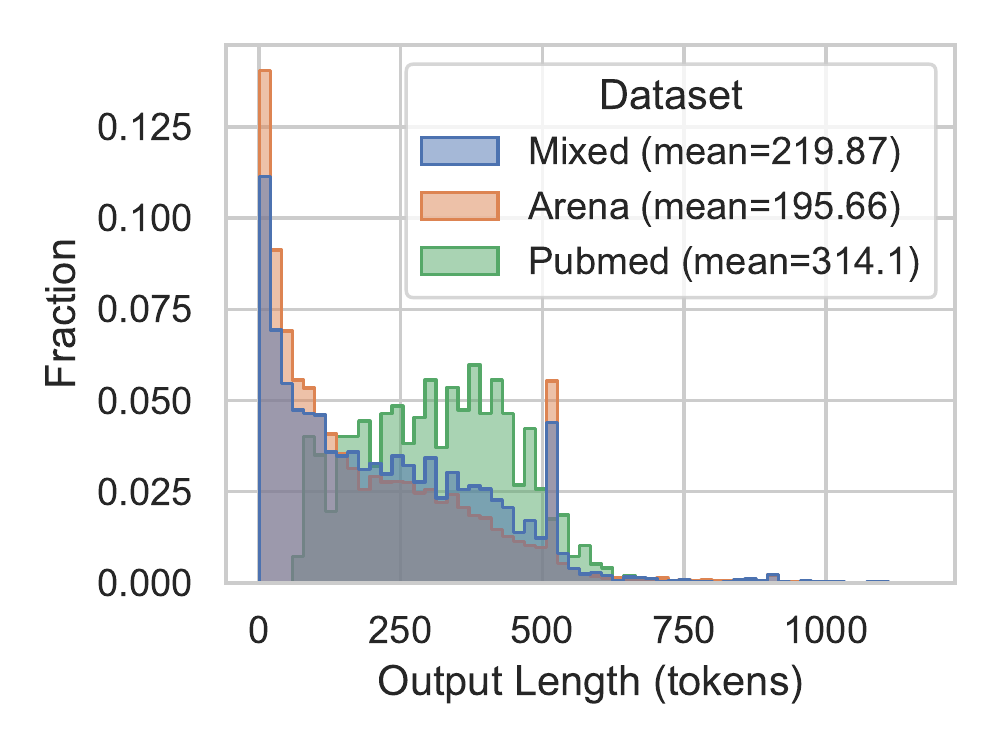}
        \captionsetup{skip=-4pt}
        \caption{Output length distributions.}
        \label{fig:data-output}
    \end{subfigure}
    % \captionsetup{skip=2pt}
    \caption{Dataset input and output length distributions.}
    \label{fig:dataset}
\end{figure}

\textbf{Datasets and SLOs.} We evaluate across three datasets to cover a wide range of application scenarios. For short-context tasks (interactive chats) we use the Chatbot Arena dataset~\citep{zheng2023judging}, for long-context tasks (document summarization) we use the PubMed dataset~\citep{Cohan_2018}, and for a mixed-context-length setting we create a synthetic dataset by sampling 80\% from Chatbot Arena and 20\% from PubMed.
The input and output length distributions are shown in ~\autoref{fig:dataset}.
We follow standard LLM inference benchmarks~\citep{llmperf-leaderboard} to set reasonable TPOT SLOs, and use 40ms to simulate services where swift responses are essential, and 120ms where longer response times are acceptable. Both selected SLOs surpass the average human reading speed, ensuring the SLOs satisfy practical user experience. 

\textbf{Mélange Configuration.}
Bucket size ranges correspond to Figure ~\ref{fig:all-gpus}, comprising of 10 input length ranges and 6 output length ranges (60 total buckets). The slice factor is set to 8 for a total of $60 \cdot 8 = 480$ slices.

\textbf{Baselines.} We compare Mélange to allocations that use a single GPU type. To derive baseline allocations, we use Mélange's ILP formulation (\S~\ref{sec:ilp}) but restrict the solver to a single GPU type. 

\subsection{Cost Savings Analysis}
\label{sec:cost-analysis}
We compare the deployment costs of Mélange to the single-GPU-type baselines across datasets and SLOs. ~\autoref{fig:cost} displays costs normalized against the cost of Mélange (purple dotted lines), and the detailed GPU allocations and cost savings are included in ~\autoref{app:allocations}.
The A10G-only and L4-only baselines are only included for the Arena dataset because the PubMed and Mixed datasets contain large requests that exceed A10G and L4's GPU memory capacity. L4 and A10G are included in Mélange's allocation but are limited to serving requests smaller than $12,000$ tokens. We now discuss each dataset in detail:

\begin{figure}[h]  
    \centering
    \includegraphics[width=0.7\linewidth]{figures/cost_legend.pdf}
    \begin{subfigure}[b]{0.33\linewidth}
        \includegraphics[width=\linewidth]{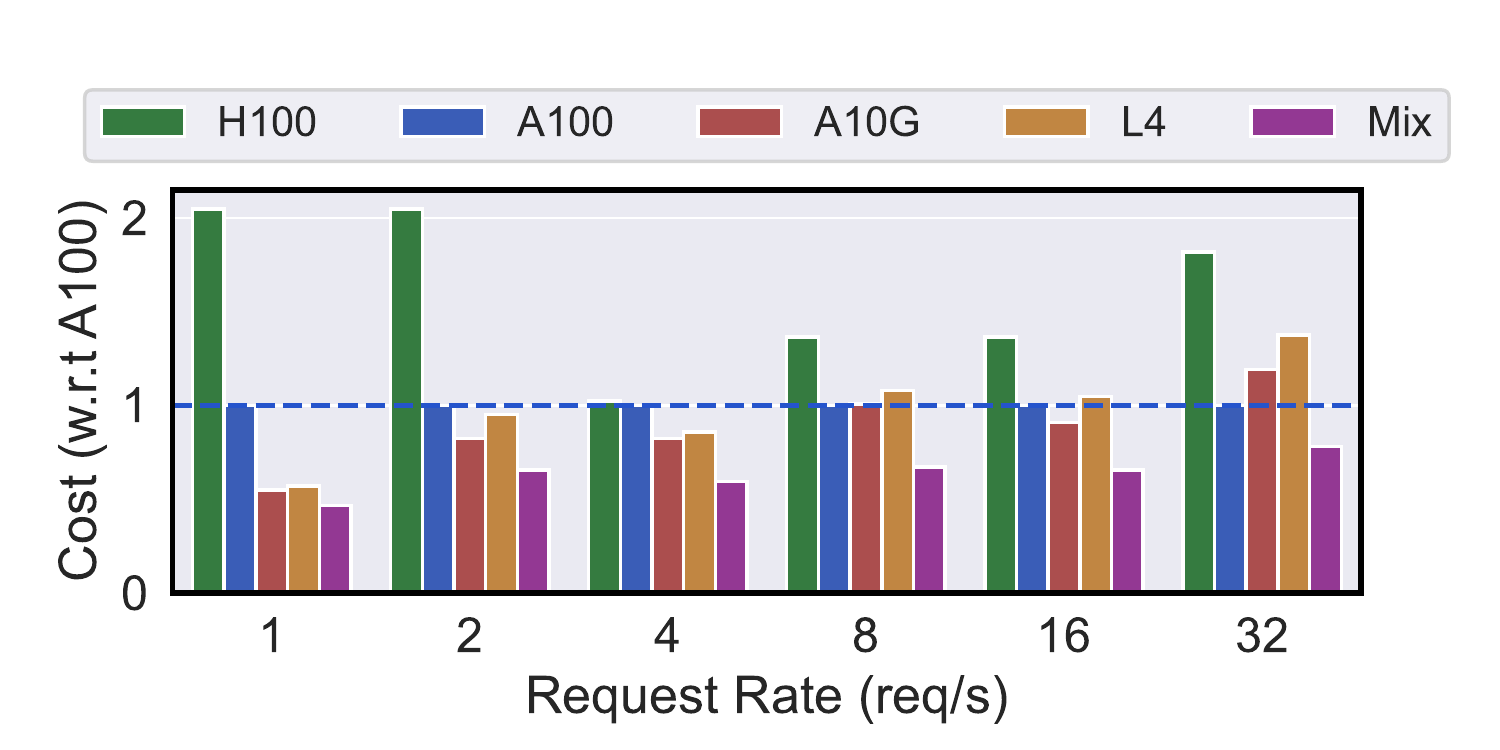}
        % \captionsetup{skip=-1pt}
        \caption{Arena, SLO = 120ms.}
        \label{fig:cost-1}
    \end{subfigure}
    \begin{subfigure}[b]{0.32\linewidth}
        \includegraphics[width=\linewidth]{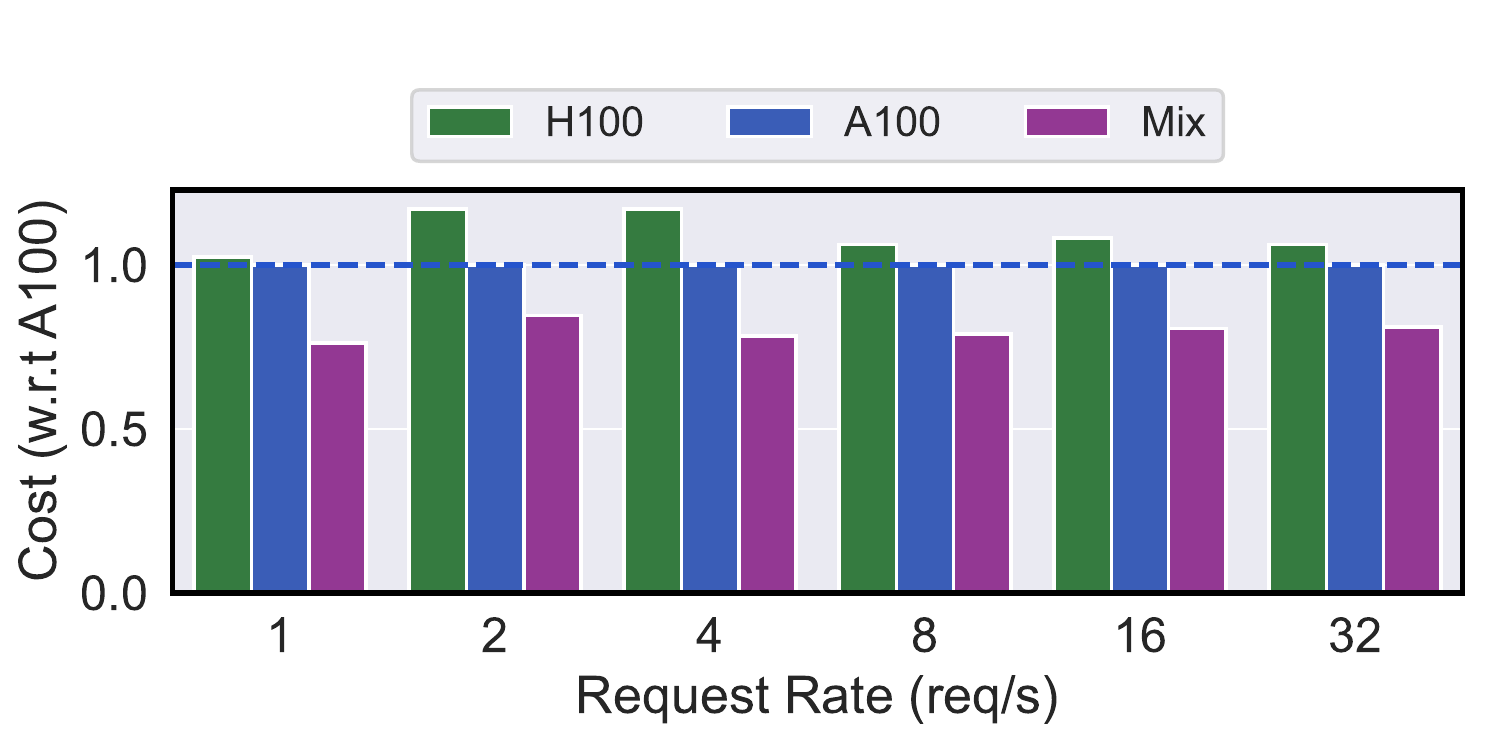}
        % \captionsetup{skip=-1pt}
        \caption{PubMed, SLO = 120ms.}
        \label{fig:cost-3}
    \end{subfigure}
    \begin{subfigure}[b]{0.32\linewidth}
        \includegraphics[width=\linewidth]{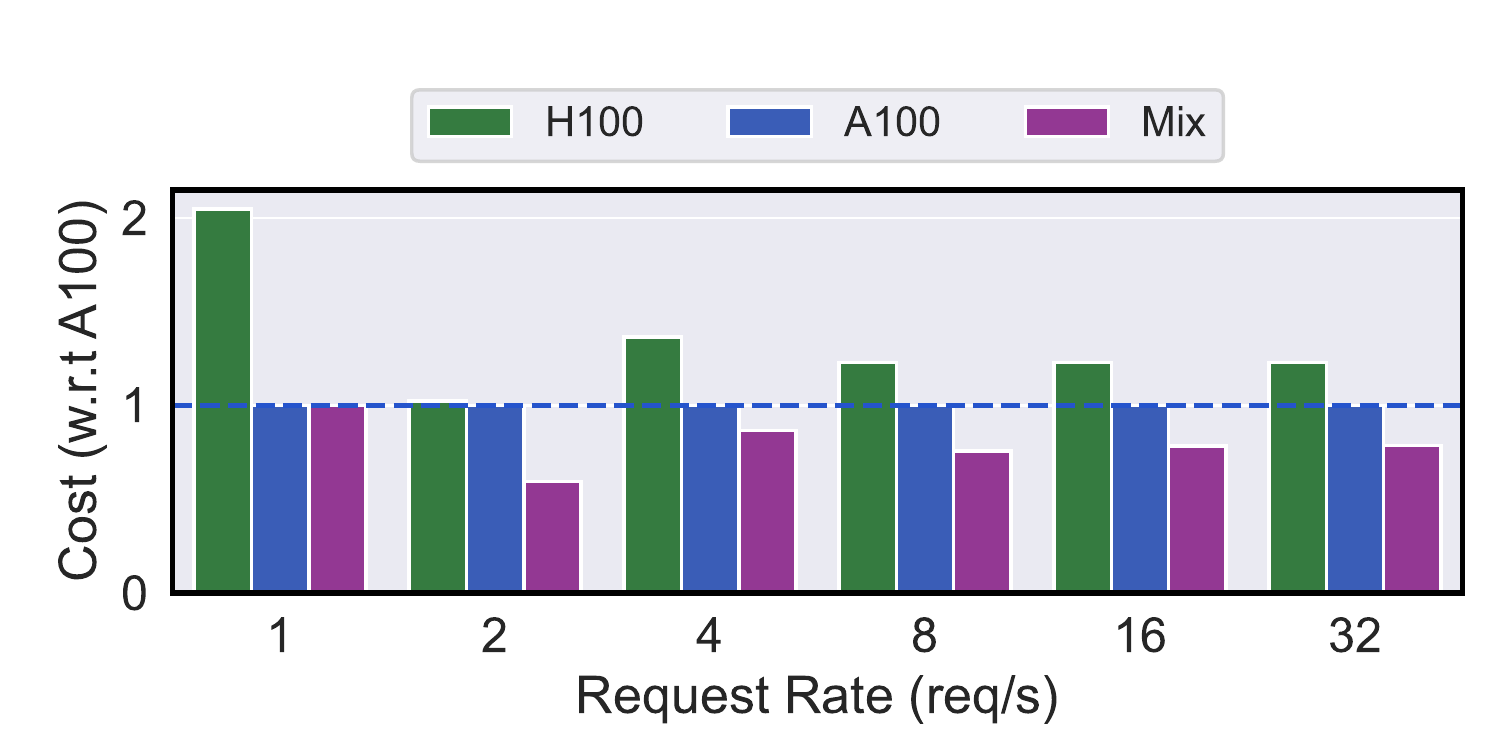}
        % \captionsetup{skip=-1pt}
        \caption{Mixed, SLO = 120ms.}
        \label{fig:cost-5}
    \end{subfigure}
    \begin{subfigure}[b]{0.33\linewidth}
        \includegraphics[width=\linewidth]{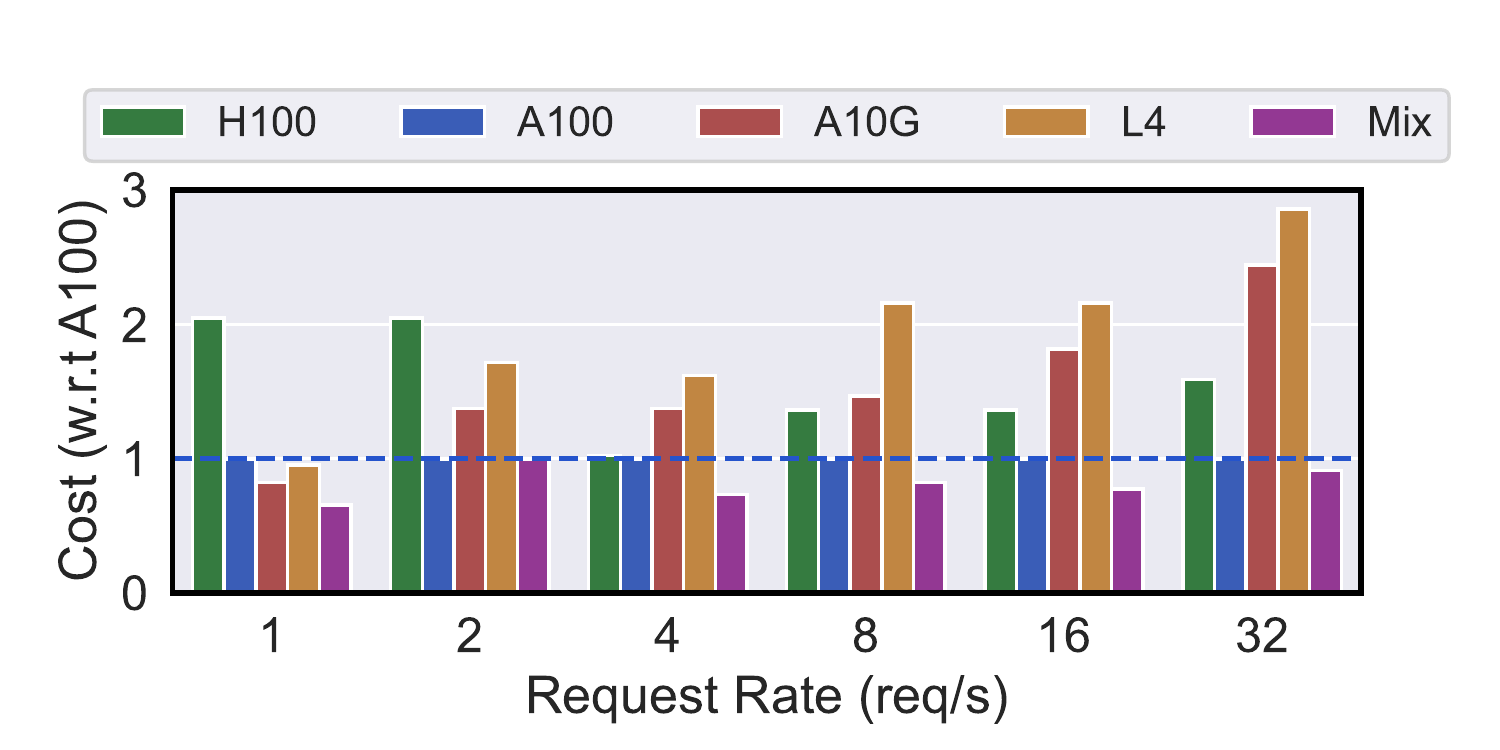}
        % \captionsetup{skip=-1pt}
        \caption{Arena, SLO = 40ms.}
        \label{fig:cost-2}
    \end{subfigure}
    \begin{subfigure}[b]{0.32\linewidth}
        \includegraphics[width=\linewidth]{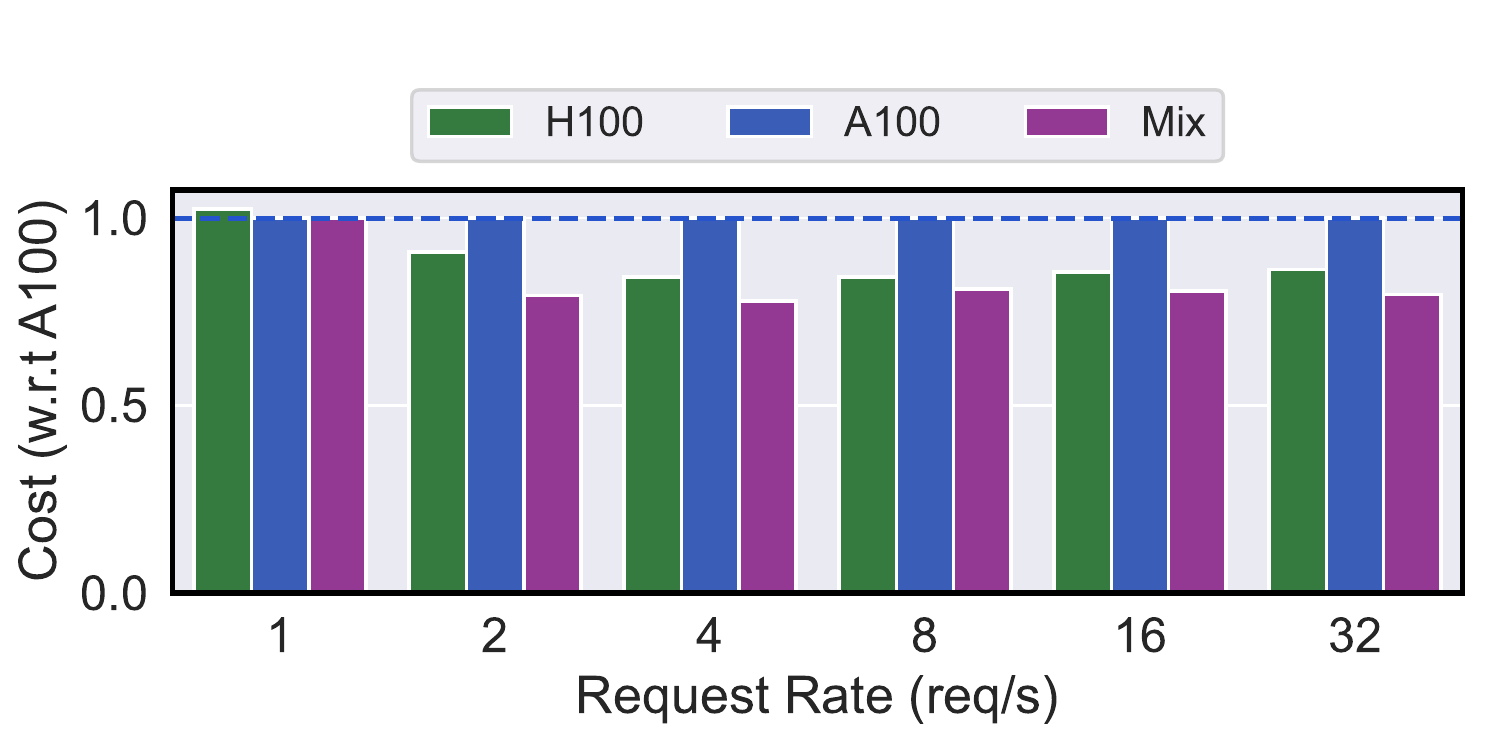}
        % \captionsetup{skip=-1pt}
        \caption{PubMed, SLO = 40ms.}
        \label{fig:cost-4}
    \end{subfigure}
    \begin{subfigure}[b]{0.32\linewidth}
        \includegraphics[width=\linewidth]{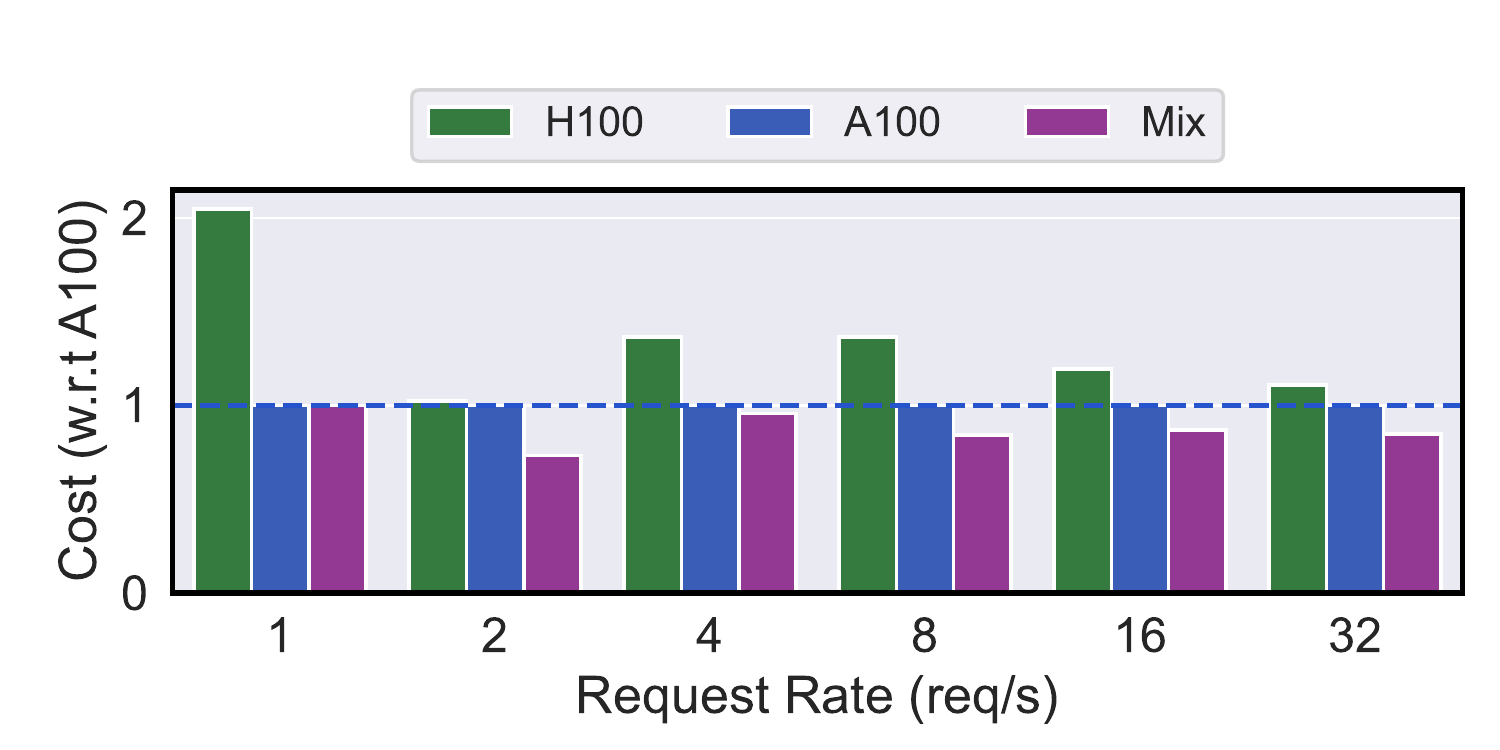}
        % \captionsetup{skip=-1pt}
        \caption{Mixed, SLO = 40ms.}
        \label{fig:cost-6}
    \end{subfigure}
    \caption{Deployment cost across different datasets and SLOs.}
    \label{fig:cost}
    % \vspace{-1.5em}
\end{figure}

\begin{itemize}[itemsep=0.8em,leftmargin=1em,rightmargin=0em]
\item \textit{Short-context Dataset (Arena).} In Figs. \ref{fig:cost-1} and \ref{fig:cost-2}, Mélange achieves 15-77\% cost reduction (120ms SLO) and 9-68\% reduction (40ms SLO). 
For both SLOs, L4/A10G are more cost efficient than A100/H100 at low request rates because they achieve greater utilization. For example, at 1-2 req/s, H100 is significantly underutilized and incurs exorbitant costs. However, as the rate increases, L4/A10G's cost advantage reduces as A100/H100 are better utilized. 
Further, with a 120ms SLO, L4/A10G remain competitive with A100 even at higher request rates due to their \tperd advantage for smaller request sizes (which the Arena dataset is skewed towards). Conversely, with a 40ms SLO, A10G/L4 show much higher relative costs due to their increased latency, requiring more instances to meet the tight deadline. Mélange adapts by allocating more L4/A10G at 120ms SLO and more A100 at 40ms SLO, consistently reducing overall cost.

\item \textit{Long-context Dataset (PubMed).} 
In Figs. \ref{fig:cost-3} and \ref{fig:cost-4}, Mélange achieves 15-33\% cost reduction (120ms SLO) and 2-22\% reduction (40ms SLO). A100 generally achieves higher \tperd for the request sizes in PubMed, evidenced by the 120ms setting where A100-only is consistently cheaper than H100-only. However, when SLO tightens to 40ms, H100 is the clear winner due to H100's lower inference latency.
Again, Mélange adapts to these dynamics by allocating a greater share of A100s at a looser SLO, and more H100s as the SLO is tightened.

\item \textit{Mixed-context Dataset.} 
In Figs. \ref{fig:cost-5} and \ref{fig:cost-6}, Mélange achieves 13-51\% cost reduction (120ms SLO) and 4-51\% reduction (40ms SLO). 
Compared to the PubMed workload, A100-only has much greater cost efficiency in the Mixed workload than H100 due to a greater portion of short-context requests, for which A100 achieves greater \tperd. Mélange capitalizes by using more A100 than H100, but it also uses L4/A10Gs for small requests, enabling even further cost reduction.
\end{itemize}

\textbf{Takeaways.} These results exemplify the core observations from \S~\ref{sec:observation}, which show that request size, SLO, and request rate all jointly determine cost efficiency. 
As any of these LLM service characteristics vary, 
Mélange flexibly adjusts its GPU allocation and mixes GPU types to exploit their heterogeneity. This consistently delivers the most cost efficient allocation across each evaluated dataset with both strict (40ms) and loose (120ms) SLOs, achieving up to a 77\% cost reduction.

\begin{wrapfigure}{r}{0.49\textwidth}
    \centering
    \includegraphics[width=\linewidth]{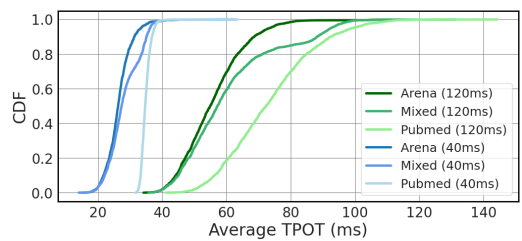}
    \caption{Mélange TPOT CDFs.}
    \label{fig:slo-cdf}
    % \vspace{-2em}
\end{wrapfigure}

\subsection{SLO Satisfaction} 
\label{sec:slo-satisfaction}
Next, we assess Mélange adherence to TPOT SLOs. We provision cloud GPU instances based on Mélange's allocation for each dataset and SLO at a rate of 4 req/s. We deploy Llama-2-7b on each GPU and sample requests randomly from the chosen dataset to serve 2K total requests. 
We record the average TPOT for each request.

\textbf{Load Balancer.} A load balancer (LB) is required to balance requests across GPUs. 
Our LB design is detailed in Appendix~\ref{app:load-balancer}. In short, the LB uses previously-served requests to estimate the output length of a new request, which is then routed to a GPU based on a weighted random selection. Weights are computed based on each GPU's performance for the request's estimated size. 

\textbf{Results.} ~\autoref{fig:slo-cdf} presents CDFs of the observed per-request average TPOTs across experiments. With an SLO of 120ms, over 99.95\% of requests met SLO. When the SLO was tightened to 40ms, 99.5\% of requests met SLO. These results validate Mélange's ability to choose GPU allocations that meet workload demand, however, we recognize that services may require even higher SLO adherence, so we investigated the source of SLO violations in our experiment.

\textbf{SLO Violation Investigation.} 84\% of our experiment's SLO violations were due to \textit{a)} request rate bursts or \textit{b)} co-location with large requests. We send requests by a Poisson process, which occasionally creates short-lived bursts that overload GPU capacity. Further, we randomly sample request sizes from the chosen dataset. Occasionally, a series of large requests are chosen in sequence and  temporarily exceed service capacity. 
In an online production environment, resource over-provisioning is used to absorb such bursts and other load variations. 
In Mélange, a desired over-provisioning rate (e.g., 10\%) can be achieved by increasing the request rate input to the solver by the same proportion.

\subsection{Solver Time}
\label{sec:solver}
We detail the solver execution time in ~\autoref{tab:solver-time}. 
Across all datasets and request rates, the solver's execution time remains under 1.2 seconds, which is negligible compared to service lifetime. We observe a modest increase in solver time with higher request volumes due to greater complexity in slice assignment.
However, this increase is empirically sub-linear relative to the increase in request rate, and the solver's execution time remains practical.

\section{Limitations and Conclusion}
\label{sec:future-work}
\textbf{Limitations.}
Mélange derives the optimal GPU allocation for a fixed workload distribution and request rate, but does not address other deployment challenges such as GPU unavailability or auto-scaling for dynamic request rates and size distributions. Mélange is only intended to make allocation decisions, a key component to be plugged into a broader serving system that handles these deployment challenges. Given the vast number of LLM deployment configurations (quantization and compression, disaggregated prefill, speculative decoding), we have not exhaustively evaluated each setting. We expect, however, that Mélange's framework is flexible to support each of these settings.

\textbf{Conclusion.} We introduce Mélange, a framework for deriving the minimal-cost GPU allocation for a given LLM service. Mélange is based on our analysis of GPU cost efficiency, which identifies three key service characteristics (request sizes, request rates, and SLOs) as significant influences on cost efficiency. 
We formulate the GPU allocation task as a cost-aware bin packing problem that accounts for each service characteristic, enabling flexibility and heterogeneity-awareness.
In evaluations on a range of GPUs, request sizes, request rates, and latency SLOs, Mélange consistently demonstrates significant reductions in deployment costs (up to 77\%) while providing high SLO attainment. 

\clearpage

\bibliographystyle{plain}
\bibliography{ref}

\appendix
\section{Experiment Setup}
\label{appendix-exp-setup}
\subsection{Dataset}
We test Mélange's performance on three different datasets listed below:
\begin{itemize}
    \item \textbf{Short context}: This scenario simulates real-time conversational dynamics by employing the Chatbot Arena dataset (\texttt{lmsys/lmsys-chat-1m})~\citep{zheng2023judging}, which is derived from real-world chatbot conversations. The dataset is skewed towards shorter context ($<2000$ tokens) because much of the data was generated in conversation with models that did not yet have a larger context window.
    \item \textbf{Long context:} This scenario represents tasks with extensive input, such as summarization. We utilize the PubMed dataset (\texttt{ccdv/pubmed-summarization})~\citep{Cohan_2018}, comprising 133 thousand scientific papers from PubMed.com, a popular dataset for large-scale text summarization studies.
    \item \textbf{Mixed long/short context:} This scenario captures settings with a combination of long and short context, such as an assistant that engages in succinct dialogue and responds to large document-based queries. To model this, we create a synthetic dataset by sampling 80\% of requests from the Arena dataset and 20\% of requests from the PubMed dataset.
\end{itemize}

\subsection{Load Balancer}
\label{app:load-balancer}
The load balancer (LB) policy used in our evaluations in \S~\ref{sec:slo-satisfaction} is as follows. For each input length bucket range (\S~\ref{sec:bucket-slice}), the LB tracks the average of all previously-seen output lengths. Upon receiving a new request, the LB uses this average as an estimate for the new request's output length, allowing the LB to identify the specific request size bucket the request belongs in. The LB then makes a weighted random selection of a GPU backend to forward the request to. A GPU's weights are computed based on the proportion of the GPU's maximum throughput for request sizes of the new request's bucket to the aggregate throughput of all GPUs. This is a simple policy we use to demonstrate the efficacy of Mélange, and leave it as future work to develop load balancers for serving LLMs on heterogeneous GPUs.

\section{Solver Time}
We present the solver execution time from each experiment in Table \ref{tab:solver-time}.

\begin{table}[!htbp]
    \centering
    \resizebox{\linewidth}{!}{
    \begin{tabular}{c|c|c|c|c|c|c}
        \toprule
        Request Rate & Arena, SLO=120ms & Arena, SLO=40ms & PubMed, SLO=120ms & PubMed, SLO=40ms & Mix, SLO=120ms & Mix, SLO=40ms  \\
        \midrule
        1 & 0.137 & 0.177 & 0.232 & 0.295 & 0.168 & 0.336 \\
        2 & 0.194 & 0.265 & 0.234 & 0.334 & 0.253 & 0.381 \\
        4 & 0.192 & 0.346 & 0.287 & 0.381 & 0.297 & 0.459 \\
        8 & 0.248 & 0.433 & 0.269 & 0.384 & 0.321 & 0.545 \\
        16 & 0.299 & 0.448 & 0.389 & 0.509 & 0.439 & 0.537 \\
        32 & 0.316 & 0.494 & 0.791 & 0.96 & 0.912 & 1.14 \\
        \bottomrule
    \end{tabular}
    }
    \caption{Solver execution time. 
    }
    \label{tab:solver-time}
\end{table}

\section{Instance Allocations} 
\label{app:allocations}

We present the instance allocations for each experiment in the tables below.

\begin{table}[!htbp]
\centering
\begin{tabular}{c|c|c|c|c|c|c|c}
\toprule
{Rate (req/s)} & {Solver} & {L4} & {A10G} & {A100} & {H100} & {\makecell{Norm. Cost \\ (\$/hr)}} & {Savings} \\ 
\midrule
1 & Mélange & 1 & 1 & & & 1.71 & {N/A} \\
  & H100-only & & & & 1 & 7.516 & 77.25\% \\
  & A100-only & & & 1 & & 3.67 & 53.41\% \\
  & A10G-only & & 2 & & & 2.02 & 15.35\% \\
  & L4-only & 3 & & & & 2.1 & 18.57\% \\
\midrule
2 & Mélange & 2 & 1 & & & 2.41 & {N/A} \\
  & H100-only & & & & 1 & 7.516 & 67.94\% \\
  & A100-only & & & 1 & & 3.67 & 34.33\% \\
  & A10G-only & & 3 & & & 3.03 & 20.46\% \\
  & L4-only & 5 & & & & 3.5 & 31.14\% \\
\midrule
4 & Mélange & 1 & & 1 & & 4.37 & {N/A} \\
  & H100-only & & & & 1 & 7.516 & 41.86\% \\
  & A100-only & & & 2 & & 7.34 & 40.46\% \\
  & A10G-only & & 6 & & & 6.06 & 27.89\% \\
  & L4-only & 9 & & & & 6.3 & 30.63\% \\
\midrule
8 & Mélange & 1 & 3 & 1 &  & 7.4 & {N/A} \\
  & H100-only & & & & 2 & 15.032 & 50.77\% \\
  & A100-only & & & 3 & & 11.01 & 32.79\% \\
  & A10G-only & & 11 & & & 11.1 & 33.39\% \\
  & L4-only & 17 & & & & 11.9 & 37.82\% \\
\midrule
16 & Mélange & 2 & 2 & 3 &  & 14.43 & {N/A} \\
   & H100-only & & & & 4 & 30.064 & 52.00\% \\
   & A100-only & & & 6 & & 22.02 & 34.47\% \\
   & A10G-only & & 20 & & & 20.2 & 28.56\% \\
   & L4-only & 33 & & & & 23.1 & 37.53\% \\
\midrule
32 & Mélange & 2 & 6 & 5 &  & 25.81 & {N/A} \\
   & H100-only & & & & 8 & 60.128 & 57.07\% \\
   & A100-only & & & 9 & & 33.03 & 21.86\% \\
   & A10G-only & & 39 & & & 39.39 & 34.48\% \\
   & L4-only & 65 & & & & 45.5 & 43.27\% \\
\bottomrule
\end{tabular}
\caption{Instance allocations for the short-context Arena dataset, SLO=120ms.}
\label{tab:allocation-arena-120ms}
\end{table}

\begin{table}[!htbp]
\centering
\begin{tabular}{c|c|c|c|c|c|c|c}
\toprule
{Rate (req/s)} & {Solver} & {L4} & {A10G} & {A100} & {H100} & {\makecell{Norm. Cost \\ (\$/hr)}} & {Savings} \\ 
\midrule
1 & Mélange &  & & 1 & 1 & 11.186 & {N/A} \\
  & H100-only & & & & 2 & 15.032 & 25.59\% \\
  & A100-Only & & & 4 & & 14.68 & 23.80\% \\
\midrule
2 & Mélange & & 3 & 1 & 2 & 21.732 & {N/A} \\
  & H100-only & & & & 4 & 30.064 & 27.71\% \\
  & A100-Only & & & 7 & & 25.69 & 15.41\% \\
  \midrule
4 & Mélange & & 3 & 4 & 3 & 40.258 & {N/A} \\
  & H100-only & & & & 8 & 60.128 & 33.05\% \\
  & A100-Only & & & 14 & & 51.38 & 21.65\% \\
  \midrule
8 & Mélange & & & 7 & 7 & 78.302 & {N/A} \\
  & H100-only & & & & 14 & 105.224 & 25.59\% \\
  & A100-Only & & & 27 & & 99.09 & 20.98\% \\
  \midrule
16 & Mélange & & & 12 & 15 & 156.78 & {N/A} \\
   & H100-only & & & & 28 & 210.448 & 25.50\% \\
   & A100-Only & & & 53 & & 194.51 & 19.40\% \\
   \midrule
32 & Mélange & 1 & 1 & 20 & 32 & 315.622 & {N/A} \\
   & H100-only & & & & 55 & 413.38 & 23.65\% \\
   & A100-Only & & & 106 & & 389.02 & 18.87\% \\
\bottomrule
\end{tabular}
\caption{Instance allocations for the long-context PubMed dataset, SLO=120ms.}
\label{tab:allocation-pubmed-120ms}
\end{table}

\begin{table}[!htbp]
\centering
\begin{tabular}{c|c|c|c|c|c|c|c}
\toprule
{Rate (req/s)} & {Solver} & {L4} & {A10G} & {A100} & {H100} & {\makecell{Norm. Cost \\ (\$/hr)}} & {Savings} \\ 
\midrule
1 & Mélange &  & & 1 & & 3.67 & {N/A} \\
  & H100-only & & & & 1 & 7.516 & 51.17\% \\
  & A100-Only & & & 1 & & 3.67 & 0\% \\
\midrule
2 & Mélange & 1 &  & 1 & & 4.37 & {N/A} \\
  & H100-only & & & & 1 & 7.516 & 41.86\% \\
  & A100-Only & & & 2 & & 7.34 & 40.46\% \\
\midrule
4 & Mélange & & 2 & & 1 & 9.536 & {N/A} \\
  & H100-only & & & & 2 & 15.032 & 36.56\% \\
  & A100-Only & & & 3 & & 11.01 & 13.39\% \\
\midrule
8 & Mélange & 1 & 2 & 1 & 1 & 13.906 & {N/A} \\
  & H100-only & & & & 3 & 22.548 & 38.33\% \\
  & A100-Only & & & 5 & & 18.35 & 24.22\% \\
\midrule
16 & Mélange & 1 & 2 & 3 & 2 & 28.762 & {N/A} \\
   & H100-only & & & & 6 & 45.096 & 36.22\% \\
   & A100-Only & & & 10 & & 36.7 & 21.63\% \\
\midrule
32 & Mélange & 1 & 5 & 6 & 4 & 57.834 & {N/A} \\
   & H100-only & & & & 12 & 90.192 & 35.88\% \\
   & A100-Only & & & 20 & & 73.4 & 21.21\% \\
\bottomrule
\end{tabular}
\caption{Instance allocations for the mixed context dataset, SLO=120ms.}
\label{tab:allocation-mixed-120ms}
\end{table}

\begin{table}[!htbp]
\centering
\begin{tabular}{c|c|c|c|c|c|c|c}
\toprule
{Rate} & {Solver} & {L4} & {A10G} & {A100} & {H100} & {\makecell{Norm. Cost \\ (\$/hr)}} & {Savings} \\ 
\midrule
1 & Mélange &  2 & 1 & & & 2.41 & N/A \\
  & H100-only & &  &  & 1 & 7.516 & 67.94\% \\
  & A100-only & &  & 1 &  & 3.67 & 34.33\% \\
  & A10G-only & & 3 & &  & 3.03 & 20.46\% \\
  & L4-only & 5 &  &  &  & 3.5 & 31.14\% \\
\midrule
2 & Mélange &  &  & 1 &  & 3.67 & N/A \\
  & H100-only & &  &  & 1 & 7.516 & 51.17\% \\
  & A100-only & &  & 1 &  & 3.67 & 0.00\% \\
  & A10G-only & & 5 &  &  & 5.05 & 27.33\% \\
  & L4-only & 9 &  &  &  & 6.3 & 41.75\% \\
\midrule
4 & Mélange & 1 & 1 & 1 &  & 5.38 & N/A \\
  & H100-only & &  &  & 1 & 7.516 & 28.42\% \\
  & A100-only & &  & 2 &  & 7.34 & 26.70\% \\
  & A10G-only & & 10 &  &  & 10.1 & 46.73\% \\
  & L4-only & 17 &  &  &  & 11.9 & 54.79\% \\
\midrule
8 & Mélange &  1 & 1 & 2 & & 9.05 & N/A \\
  & H100-only & &  &  & 3 & 15.032 & 39.80\% \\
  & A100-only & &  & 3 &  & 11.01 & 17.80\% \\
  & A10G-only & & 16 &  &  & 16.16 & 44.00\% \\
  & L4-only & 34 &  &  &  & 23.8 & 61.97\% \\
\midrule
16 & Mélange &  & 6 & 3 & & 17.07 & N/A \\
   & H100-only & &  &  & 4 & 30.064 & 43.22\% \\
   & A100-only & &  & 6 &  & 22.02 & 22.48\% \\
   & A10G-only & & 40 &  &  & 40.4 & 57.75\% \\
   & L4-only & 68 &  &  &  & 47.6 & 64.14\% \\
\midrule
32 & Mélange &  & 8 & 6 &  & 30.1 & N/A \\
   & H100-only & &  &  & 7 & 52.612 & 42.79\% \\
   & A100-only & &  & 9 &  & 33.03 & 8.87\% \\
   & A10G-only & & 80 &  &  & 80.8 & 62.75\% \\
   & L4-only & 135 &  &  &  & 94.5 & 68.15\% \\
\bottomrule
\end{tabular}
\caption{Instance allocations for the short-context Arena dataset, SLO=40ms.}
\label{tab:allocation-arena-40ms}
\end{table}

\begin{table}[!htbp]
\centering
\begin{tabular}{c|c|c|c|c|c|c|c}
\toprule
{Rate (req/s)} & {Solver} & {L4} & {A10G} & {A100} & {H100} & {\makecell{Norm. Cost \\ (\$/hr)}} & {Savings} \\ 
\midrule
1 & Mélange &  &  & 4 &  & 14.68 & N/A \\
  & H100-only & & &  & 2 & 15.032 & 2.34\% \\
  & A100-Only & & & 4 &  & 14.68 & 0.00\% \\
\midrule
2 & Mélange &  & & 1 & 3  & 26.218 & N/A \\
  & H100-only & & &  & 4 & 30.064 & 12.79\% \\
  & A100-Only & & & 9 &  & 33.03 & 20.62\% \\
\midrule
4 & Mélange & & & 3 & 5  & 48.59 & N/A \\
  & H100-only & & &  & 7 & 52.612 & 7.64\% \\
  & A100-Only & & & 17 &  & 62.39 & 22.12\% \\
\midrule
8 & Mélange & & & 3 & 12 & 101.202 & N/A \\
  & H100-only & & &  & 14 & 105.224 & 3.82\% \\
  & A100-Only & & & 34 &  & 124.78 & 18.90\% \\
\midrule
16 & Mélange & & & 11 & 21 &  198.206 & N/A \\
   & H100-only & & &  & 28 & 210.448 & 5.82\% \\
   & A100-Only & & & 67 &  & 245.89 & 19.39\% \\
\midrule
32 & Mélange & & & 24 & 40 &  388.72 & N/A \\
   & H100-only & & &  & 56 & 420.896 & 7.64\% \\
   & A100-Only & & & 133 &  & 488.11 & 20.36\% \\
\bottomrule
\end{tabular}
\caption{Instance allocations for the long-context PubMed dataset, SLO=40ms.}
\label{tab:allocation-pubmed-40ms}
\end{table}

\begin{table}[!htbp]
\centering
\begin{tabular}{c|c|c|c|c|c|c|c}
\toprule
{Rate (req/s)} & {Solver} & {L4} & {A10G} & {A100} & {H100} & {\makecell{Norm. Cost \\ (\$/hr)}} & {Savings} \\ 
\midrule
1 & Mélange &  &  &  1 & & 3.67 & N/A \\
  & H100-only & & &  & 1 & 7.516 & 51.17\% \\
  & A100-only & & & 1 &  & 3.67 & 0.00\% \\
\midrule
2 & Mélange & 1 & 1 & 1 &  & 5.38 & N/A \\
  & H100-only & & &  & 1 & 7.516 & 28.42\% \\
  & A100-only & & & 2 &  & 7.34 & 26.70\% \\
\midrule
4 & Mélange &  & 3 &  & 1 & 10.546 & N/A \\
  & H100-only & & &  & 2 & 15.032 & 29.84\% \\
  & A100-only & & & 3 &  & 11.01 & 4.21\% \\
\midrule
8 & Mélange &  1 & 3 & 2 & 1 & 18.586 & N/A \\
  & H100-only & & &  & 4 & 30.064 & 38.18\% \\
  & A100-only & & & 6 &  & 22.02 & 15.59\% \\
\midrule
16 & Mélange & 2 & 7 & 2 & 3 & 38.358 & N/A \\
   & H100-only & & &  & 7 & 52.612 & 27.09\% \\
   & A100-only & & & 12 &  & 44.04 & 12.90\% \\
\midrule
32 & Mélange &  & 15 & 6 & 5 & 74.75 & N/A \\
   & H100-only & & &  & 13 & 97.708 & 23.50\% \\
   & A100-only & & & 24 &  & 88.08 & 15.13\% \\
\bottomrule
\end{tabular}
\caption{Instance allocations for the mixed long/short context dataset, SLO=40ms.}
\label{tab:allocation-mixed-40ms}
\end{table}

\clearpage
\end{document}